\documentclass[nofootinbib,preprint,superscriptaddress,prA]{revtex4-2}
\pdfoutput=1

\usepackage{multirow}
\usepackage{siunitx}
\usepackage{tikz}
\usetikzlibrary{patterns}
\usetikzlibrary{decorations.pathmorphing}
\usepackage[scr=boondoxo]{mathalpha}
\newcommand{\lfont}{\fontfamily{qzc}\selectfont}
\usepackage{graphicx,times}
\usepackage{latexsym}
\usepackage{physics}
\usepackage{mathtools}
\usepackage{amsmath,amssymb,amsbsy,amsfonts}
\usepackage[nointegrals]{wasysym}
\usepackage{array}
\usepackage{graphics}
\usepackage{xcolor}
\usepackage{cancel}
\usepackage{xspace}
\usepackage[normalem]{ulem}
\usepackage[unicode, pdfprintscaling=None]{hyperref}
\hypersetup{
    colorlinks,
    linkcolor={red!50!black},
    citecolor={blue!50!black},
    urlcolor={blue!80!black}
}

\usepackage[capitalise]{cleveref}

\usepackage{xcolor}
\usepackage{makecell}

\newcommand{\vect}[1]{\vec{\boldsymbol{#1}}}
\newcommand{\hatvect}[1]{\widehat{\boldsymbol{#1}}}

\newcommand{\CG}[6]{C_{#1,#2,#3}^{#4,#5,#6}}

\newcommand{\eftnopi}{EFT($\cancel{\pi}$)\xspace}

\usepackage{orcidlink}

\usepackage{relsize}

\usepackage{etoolbox}

\makeatletter
\newcommand*{\rom}[1]{\expandafter\@slowromancap\romannumeral #1@}

\newcommand{\JacobiPW}{a partial wave basis of Jacobi momenta}
\newcommand{\fourBodyMom}{single-particle momenta }
\newcommand{\fourBodyMomdash}{single-particle momentum }
\newcommand{\NormPW}{a partial wave basis of \fourBodyMom}
\begin{document}
\title{Four-Body Systems at Large Cutoffs in Effective Field Theory}

\author{Xincheng Lin\,\orcidlink{0000-0001-9068-6787}}
\email{xincheng.lin@duke.edu}
\affiliation{Department of Physics, Duke University, Durham, North Carolina 27708, USA}
\begin{abstract}
    Four-body systems are studied using an effective field theory with two- and three-body contact interactions.
    A method to systematically address deep trimers (three-body bound states that are more tightly bound than four-body bound/resonant states) in four-body calculations is developed using a diagrammatic approach.
    Previous calculations were limited by the existence of deep trimers, which this work overcomes.
    For cold $^4$He atoms, binding energies of 526.1(5)~mK and 128.517(1)~mK are obtained at leading order for the tetramer ground and excited states, respectively, where errors come from the truncation of three-body partial waves. Tetramer binding energies and decay widths are also computed approaching the unitary limit. In the unitary limit, there are two tetramers associated with each trimer of binding energy $B_3^{(0)}$. The binding energy and decay width for the associated tetramer ground state are $E_4^{(0)} = 4.60(1)B_3^{(0)}$ and $\Gamma_4^{(0)}/2 = 0.0160(1)B_3^{(0)}$, respectively, and for the associated tetramer excited state, $E_4^{(1)} = 1.0022(3)B_3^{(0)}$ and $\Gamma_4^{(1)}/2 = 2.57(2)\times 10^{-4}B_3^{(0)}$, respectively.
    This calculation is a gateway to higher-order and/or more-body calculations in nuclear and atomic systems.
\end{abstract}
\maketitle
\newpage
\section{Introduction}
Quantum systems with a typical scale, $a_{\textrm{sys}}$, that is much larger than the range of their underlying interaction, $l_{\textrm{sys}}$, exist in both atomic and nuclear systems, such as cold $^4$He atoms, few-nucleon systems, and nuclear halo systems. These systems share many universal features due to their proximity to the unitary limit and can be systematically described using effective field theory (EFT). 
An EFT includes all possible interactions allowed by the underlying symmetries. The interactions are ordered by the EFT power counting with an expansion parameter given by, for example,  the ratio $l_{\textrm{sys}}/a_{\textrm{sys}}$. Such power counting is usually obtained using naive dimensional analysis (NDA) of the interactions informed by the requirement that the observables must be renormalization group (RG) invariant; the latter can be accomplished by studying the regularization dependence, e.g., loop integral dependence on a cutoff.\footnote{The cutoff used in the loop integrals and the cutoff of the EFT are different objects. The former will be referred to as ``cutoff'' in this paper unless specified otherwise.}Because of its power counting, an EFT has the advantage of giving a model-independent prediction of observables with a systematic error estimation at each order.

Non-relativistic short-range EFTs have found great success in describing few-body nuclear and atomic systems after the power counting and regularization in such EFTs became understood~\cite{Kaplan_1998_1, Kaplan_1998_2}. Universal features have been found that lead to a deeper understanding of few-body systems~\cite{Braaten_2006}.
Three-boson systems with short-range interactions were studied by Bedaque et al.~\cite{Bedaque_1999,Bedaque_1999b} using an EFT approach, who computed the dimer (two-boson bound state)-single-boson scattering amplitude diagrammatically. While diagrams for the LO scattering amplitude only consist of two-boson interactions according to NDA, the scattering amplitude is not convergent as a function of cutoff. In order to maintain the RG invariance of the scattering amplitude, a three-boson contact interaction is promoted to LO and is fit to a three-boson observable, e.g. a trimer (three-boson bound state) binding energy, at each cutoff. Moreover, as the cutoff increases, deeply bound trimer states, known as Efimov trimers, appear. These trimers correspond to poles in the dimer-single-boson scattering amplitude, even though their binding momenta may be larger than the cutoff of the effective theory. The need for a three-body contact interaction carries over to three-nucleon systems in order to renormalize the LO spin-doublet ($S = 1/2$) channel~\cite{Bedaque_2000}. In fact, the triton can be understood as an Efimov trimer of one proton and two neutrons~\cite{Hammer_2010, Hammer_2020}.

Four-boson systems with a large two-body scattering length were studied by Platter et al.~\cite{Platter:2004he} using an effective theory approach, where the Faddeev-Yakubovsky (FY) equation~\cite{Yakubovsky:1966ue} was solved with effective potentials that have the form of a delta function at LO (and derivatives of delta functions at higher order), taken from the two- and three-boson contact interactions in EFT. They considered cold $^4$He atoms and computed the trimer and tetramer (four-boson bound state) binding energies as a function of cutoff. Their calculation, however, only works below a threshold cutoff $\Lambda_t$, above which a deeply bound trimer appears and creates instabilities in four-boson calculations. Above the threshold cutoff, the tetramers become resonances with an open decay channel into this deeply bound trimer. Nevertheless, they argued that the tetramer binding energies appear to be converging for cutoffs below $\Lambda_t$ and estimated their errors from the residual cutoff dependence in addition to errors from the EFT expansion. Their results were in good agreement with Blume and Greene~\cite{Blume2000MonteCH}, who used the LM2M2 potential~\cite{LM2M2} and combined Monte
Carlo methods with the adiabatic hyperspherical approximation. Platter et al.~\cite{Platter:2004he} concluded that there is no need for a four-boson contact interaction at LO to renormalize the four-boson system. A four-boson contact interaction was later found by Bazak et al.~\cite{Bazak_2019} to be required at next-to-leading order (NLO), where the two-body effective range correction is included, in order to maintain RG invariance of the tetramer ground state binding energy. Another four-boson calculation of cold $^4$He atoms using the FY equation with the LM2M2 potential can be found in Ref.~\cite{Lazauskas_2006}. Universality in four-boson systems was studied by Hammer and Platter~\cite{Hammer_2007} using two- and three-body contact interactions. They conjectured that there exist two (resonantly bound) tetramer states with binding energies located between the binding energies of any two adjacent Efimov trimers due to the discrete scaling symmetry. This conjecture was supported by von Stecher et al.~\cite{2009NatPh...5..417V}, who studied the four-boson system in the unitary limit, where the two-boson scattering length, $a$, approaches infinity. This was also investigated theoretically by Deltuva~\cite{Deltuva:2010xd} and Hadizadeh et al.~\cite{Hadizadeh2011}. More recently, Frederico and Gattobigio~\cite{frederico2023universal} studied the tetramer limit cycle in the unitary limit by tuning a four-body scale.

The study of four-boson systems was extended to four-nucleon systems by Platter et al.~\cite{Platter_2005}, who used the FY equation with two- and three-nucleon contact interactions taken from LO pionless EFT (\eftnopi).  The binding energy of the $\alpha$-particle was computed and found to scale approximately linearly with the triton binding energy, in agreement with the well-known Tjon line~\cite{Tjon:1975sme}. For the cutoffs considered in their calculation, there is only one three-nucleon bound state, i.e., no deeply bound Efimov state exists. Similar to the four-boson case, they argued that no four-nucleon force is required at LO to renormalize the four-nucleon system. Nucleon-trinucleon scatterings were investigated using the resonating group method with \eftnopi by Kirscher et al. ~\cite{Kirscher_2010} and Kirscher~\cite{Kirscher:2011uc}, who also discussed the cutoff dependence of their results.

An alternative approach to treating four-body systems is the diagrammatic approach~\cite{PhysRevA.73.032724}, where the four-body integral equations are derived from Feynman diagrams. The diagrammatic four-boson integral equation in two and three spatial dimensions (2D and 3D, respectively) was solved by Brodsky et al.~\cite{PhysRevA.73.032724}. However, no explicit three-boson force was included, and the cutoff dependence of the results was not discussed. It is also noteworthy that, unlike three-body systems where each Feynman diagram can be matched to a term in the Faddeev equation, for four-body systems it is unclear how to build a one-to-one correspondence between each term in the diagrammatic and FY equations (see, e.g., Ref.~\cite{Blokhintsev:1983vv}). For four- and/or more-boson calculations using other methods, see Refs.~\cite{2009NatPh...5..417V, von_Stecher_2010, Hiyama_2012, Gattobigio_2012,Bazak:2016wxm}.
 
The main purpose of this paper is to investigate the cutoff dependence of tetramer binding energies at cutoffs above $\Lambda_t$, where one or more deep trimers exist, in four-boson systems with a large two-body scattering length. 
This LO four-body calculation at large cutoffs is a first step to higher-order and/or more-body calculations in nuclear and atomic systems. Using a full EFT approach, the four-boson integral equation is first obtained from the Feynman diagrams with two- and three-boson contact interactions and then rewritten in terms of the three-boson amplitudes. This facilitates the inclusion or subtraction of the deep trimers in the four-body calculation. 
For physical systems where the deep trimers exist, it is necessary to include them in four-body calculations in order to obtain the binding energy and decay width of the tetramer resonances above these deep trimers. For physical systems where the deep trimers do not exist, such as cold $^4$He atoms, it is necessary to subtract them from the four-body calculation in order to compute observables at cutoffs above $\Lambda_t$.

As an example of the four-boson calculation, tetramer binding energies for cold $^4$He atoms are computed at cutoffs above $\Lambda_t$ in this study. The convergence of the diagrammatic results at sufficiently large cutoffs is demonstrated. This convergence at large cutoffs is important to NLO or higher-order calculations. This is because for cold $^4$He atoms the EFT error at NLO is only $\approx 1\%$ using an EFT expansion parameter of $\approx 10\%$~\cite{Platter:2004he}, while the tetramer binding energies obtained using cutoffs below $\Lambda_t$ contain a $\approx 5\%$ ($\approx 2\%$) error from the residual cutoff dependence for the tetramer ground (excited) state~\cite{Platter:2004he}. 
This means errors from the residual cutoff dependence will obscure or even dominate the EFT error in an NLO or higher-order calculation if cutoffs are limited below $\Lambda_t$. A similar situation happens to the four-nucleon system, where the $\alpha$-particle binding energy obtained by Platter et al.~\cite{Platter_2005} has a $\approx 5\%$ variation in the range of their cutoff between $8$~fm$^{-1}$ and $10$~fm$^{-1}$. This variation is small compared to the LO \eftnopi error of $\approx 30\%$ but sizable compared to the NLO \eftnopi error of $\approx 10\%$. 
Therefore, the convergence of four-body observables as functions of cutoff is necessary for minimizing uncertainties at NLO or higher order. In addition, the convergence at cutoffs above $\Lambda_t$ reaffirms that no four-body force is required at LO. This is important since, as demonstrated later by the cutoff dependence of the tetramer binding energies, the results obtained with cutoffs below $\Lambda_t$ may be misleading.\footnote{Precedent in nuclear systems exists where the need for a counterterm at large cutoffs can be obscured by results obtained with relatively low cutoffs~\cite{coulombconf}.}

This paper is organized as follows. In Sec.~\ref{sec:3bosonSys} two- and three-boson systems in an EFT with two- and three-boson contact interactions are reviewed and the three-boson amplitude and trimer residues are calculated. Sec.~\ref{sec:4bodySys-formal} shows the diagrammatic and operator forms of the four-boson integral equation. In Sec.~\ref{sec:4bodySys-expressions} this equation is projected onto a partial-wave basis and its kernel is expressed in terms of three-boson amplitudes to include the trimer poles. Sec.~\ref{sec:4bodySys-implementation} discusses how to implement the diagrammatic approach to cold $^4$He atoms and calculate tetramer binding energies, and how to approach the unitary limit in this calculation. Sec.~\ref{sec:results} demonstrates the convergence of the tetramer binding energies for cold $^4$He atoms as a function of cutoff and the results in the unitary limit. Sec.~\ref{sec:conclusions} summarizes this work and discusses future directions. Appendices include further details of the four-boson calculations.

\section{Two- And Three-Boson Systems}
\label{sec:3bosonSys}
\begin{figure}[htb]
    \centering
    \includegraphics[width=\textwidth]{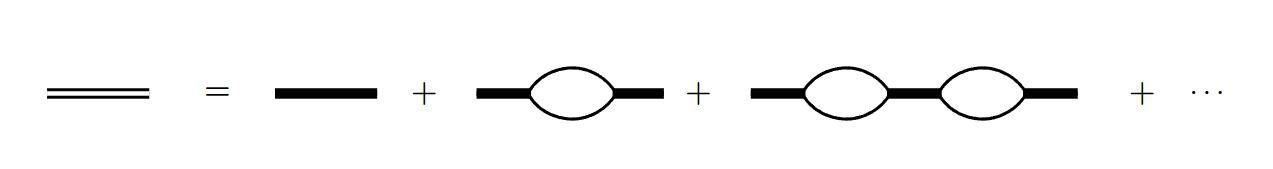}
    \caption{This equation gives the LO dressed dimer propagator, represented by the double lines on the left-hand side. The solid bar and the single lines on the right-hand side are the bare dimer and single boson propagators, respectively. The right-hand side is a geometric sum that can be computed analytically.}
    \label{fig:dressed dimer}
\end{figure}
For a non-relativistic boson system, the leading-order (LO) two-body Lagrangian is
\begin{align}
    \mathcal{L}_2 = \psi^\dagger(i\partial_0 + \frac{\vect{\nabla}^2}{2m})\psi + d^\dagger \Delta d + \frac{y}{\sqrt{2}}(d^\dagger \psi\psi + \textrm{h.c.}),
\end{align}
where $\psi$ is the single-boson field with mass $m$ and $d$ is the dimer auxiliary field. The dimer auxiliary field can be integrated out and the only free parameter in this two-body Lagrangian is $y^2/\Delta$ (see, e.g., Ref.~\cite{Bedaque_1999}).
The dressed dimer propagator is given by the geometric sum (also shown diagrammatically in Fig.~\ref{fig:dressed dimer}):
\begin{align}
    iD_{d}(p_0, \vect{{p}}) = \frac{i}{\Delta - \frac{y^2m}{4\pi}(-\mu + \sqrt{\frac{\vect{{p}}^2}{4}-mp_0 - i\epsilon})}, 
    \label{eq:dimerprop}
\end{align}
where $\mu$ is a non-physical scale emerging from the linear divergence of the loop integral in 3D, coming from dimensional regularization with the power divergence subtraction \cite{Kaplan_1998_1, Kaplan_1998_2}. The combination $y^2/\Delta$ is obtained by fitting Eq.~\eqref{eq:dimerprop} to the physical dimer pole and choose the following parameterization for $y^2$ and $\Delta$
\begin{align}
\centering
y^2 = \frac{4\pi}{m},\qquad 
\Delta = \gamma - \mu, 
\end{align}
where $\gamma$ is the dimer binding momentum. The dressed dimer propagator becomes
\begin{align}
    iD_{d}(p_0, \vect{{p}}) = 
    \frac{i}{\gamma - \sqrt{\frac{\vect{{p}}^2}{4}-mp_0 - i\epsilon}},
\end{align}
which does not have any regulator dependence. This means, as discussed in Ref.~\cite{Bedaque_2000}, that the cutoff of the two-boson system is taken to infinity before the cutoff of three- and four-boson systems, in contrast to taking the cutoff of the two-, three-, and four-boson systems to infinity at the same time as was done in, e.g., Ref.~\cite{Platter:2004he}.

One way to write the three-boson Lagrangian is
\begin{align}
    \mathcal{L}_3 = h\psi^\dagger d^\dagger \psi d,
    \label{eqn:three-boson force}
\end{align}
where $h$ is the dimer-single-boson coupling coefficient.
A more convenient approach, as used in the rest of this paper, is to introduce a trimer auxiliary field, $\tau$, and write the three-boson Lagrangian as
\begin{align}
    \mathcal{L}'_3 = \tau^\dagger \Omega \tau + \omega (\tau^\dagger \psi d + \textrm{h.c.}).
    \label{eq:L3aux}
\end{align}
Integrating out the trimer auxiliary field gives the matching condition
\begin{align}
\centering
-\frac{\omega^2}{\Omega} = h.
\label{eqn:hmatching}
\end{align}
The only free parameter in the three-body Lagrangian is $h$, whose size can be determined by fitting to three-boson observables, such as the dimer-single-boson scattering length or trimer binding energy. 
\begin{figure}[htb]
    \centering    
    \includegraphics[width=\textwidth]{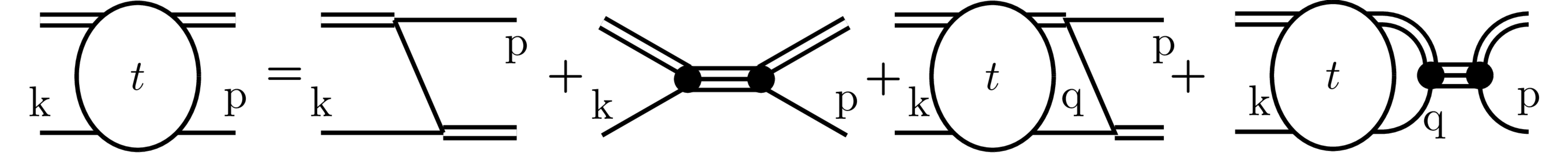}
    \caption{Diagrammatic representation of the integral equation for the dimer-single-boson scattering amplitude, $t$. Triple line represents bare trimer propagator, $i/\Delta$, and solid dot represents three-boson coupling, $i\omega$. $\textrm{k}$ ($\textrm{p}$) is the on-shell four-momentum of the incoming (outgoing) single boson in the CM frame. $\textrm{q}$ is that for the internal single boson.}
    \label{fig:dimer boson intg}
\end{figure}

Fig.~\ref{fig:dimer boson intg} shows diagrammatically the integral equation for the dimer-single-boson scattering amplitude~\cite{Bedaque_1999,Bedaque_1999b}, whose operator form is
\begin{align}
    t = M + Kt,
    \label{eq:3boson-OpForm}
\end{align}
where $t$ is the dimer-single-boson scattering amplitude (also referred to as the three-boson amplitude), $M$ is the inhomogeneous term, and $K$ is the kernel of the integral equation. To project this equation onto a momentum basis, one can consider the four-momentum of the single boson legs in the center-of-mass (CM) frame, as indicated in Fig.~\ref{fig:dimer boson intg}. For the loop energy integrals, one can invoke the residue theorem and pick up the single-boson poles. Upon discretization one finds a closed system of equations corresponding to the integral equation in Eq.~\eqref{eq:3boson-OpForm} with single-boson momenta, $\textrm{p}$, $\textrm{q}$, and $\textrm{k}$ being on-shell, i.e., $\textrm{p} = \{p^2/(2m), \vect{p}\}$, $\textrm{q} = \{q^2/(2m), \vect{q}\}$, and $\textrm{k} = \{k^2/(2m), \vect{k}\}$. Eq.~\eqref{eq:3boson-OpForm} separates under a partial-wave basis, and the $\ell$-th 
 partial wave gives~\cite{Bedaque_1999,Bedaque_1999b,Gabbiani:1999yv}
\begin{align}
    t^h_{\ell}(E, k, p) = M_{\ell}^h(E,k,p) + \int_{0}^{\Lambda} \frac{q^2dq}{2\pi^2} t^h_{\ell}(E, k, q)K_{\ell}^h(E,q,p),
\label{eqn:integral equation for three-boson scatterings}
\end{align}
where $E$ is the total energy in the three-boson CM frame and $t^h_{\ell}(E, k, p)$ is the matrix element of $t$ in the $\ell$-th partial wave channel with an incoming (outgoing) single-boson on-shell three-momentum $k$($p$). The superscript $h$ indicates the inclusion of the three-boson force. A sharp cutoff $\Lambda$ is used in three-boson systems. 
$M_{\ell}^h(E,k,p)$ and $K_{\ell}^h(E,q,p)$ are the matrix elements of $M$ and $K$, respectively, and read
\begin{align}
    \centering
   M_{\ell}^h(E,k,p) &= \frac{8\pi}{kp}Q_{\ell}\left(\frac{k^2 + p^2 - mE -i\epsilon}{kp}\right) + h\delta_{\ell0} \label{eq:B_l^h}\\
   K_{\ell}^h(E,q,p) &= -D_d(E-\frac{q^2}{2m}, \vect{q})\left(\frac{8\pi}{kp}Q_{\ell}\left(\frac{q^2 + p^2 - mE -i\epsilon}{kp}\right) + h\delta_{\ell0}\right),
   \label{eq:K_l^h}
\end{align}
where the matching for the three-boson force in Eq.~\eqref{eqn:hmatching} is used. The $Q_{\ell}(x)$ are the Legendre functions of the second
kind defined as
\begin{align}
    Q_{\ell}(x) \equiv \int_{-1}^{1}\frac{du}{2}\frac{P_{\ell}(u)}{u+x},
    \label{eq:Qlegendre}
\end{align}
where $P_{\ell}(u)$ are the Legendre polynomials. Note that Eq.~\eqref{eq:Qlegendre} differs from the conventional definition of the Legendre functions of the second
kind by a phase of $(-1)^{\ell}$.

Unlike two-body scattering whose analytic solution is known, Eq.~\eqref{eqn:integral equation for three-boson scatterings} has only been solved numerically. While NDA suggests that the three-boson operator, which has a higher mass dimension, should not be included at LO, a careful study \cite{Bedaque_1999} of this integral equation shows that without this three-boson interaction, there exists a zero mode that leads to cutoff dependence of the solution; in other words, the renormalization of this three-boson problem requires this three-boson interaction to appear at LO. Higher order corrections to this three-boson force can be included perturbatively and energy-dependent terms are needed starting at next-to-next-to-leading order (NNLO) \cite{BEDAQUE2003589,Ji_2013}. The sizes of the three-boson forces at each order can be determined at each cutoff by fitting to three-boson observables, such as a trimer binding energy or dimer-single-boson scattering length. 
In addition, an infinite number of three-boson bound states occur as the cutoff goes to infinity. When $E$ equals the binding energy of each trimer, the kernel matrix of the (discretized) integral equation in Eq.~\eqref{eqn:integral equation for three-boson scatterings} has an  eigenvalue equal to one. 

\begin{figure}[htb]
    \centering    
    \includegraphics[width=6in]{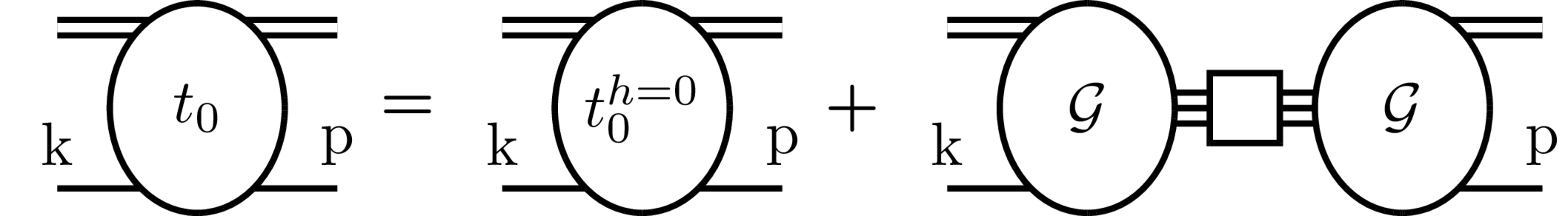}
    \caption{S-wave three-boson scattering amplitude rewritten in terms of the trimer vertex function and the dressed trimer propagator.}
    \label{fig:dimer-boson-rewritten}
\end{figure}

\begin{figure}[htb]
    \centering    
    \includegraphics[width=5in]{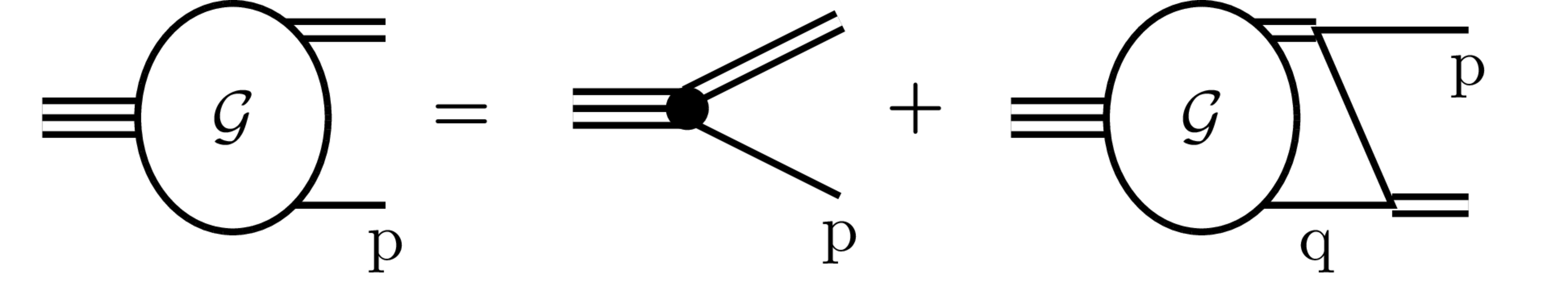}
    \caption{Integral equation for the trimer vertex function $\mathcal{G}$.}
    \label{fig:trimer vertex function}
\end{figure}

\begin{figure}[htb]
    \centering    
    \includegraphics[width=\textwidth]{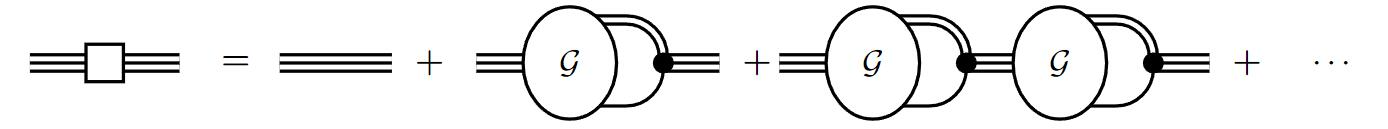}
    \caption{Equation for LO dressed trimer propagator, represented by the triple lines with a box on the left-hand side. The right-hand side is a geometric sum that can be expressed in $\Sigma_{\Lambda}(E)$.}
    \label{fig:dressed trimer}
\end{figure}

Expanding the S-wave three-boson scattering amplitude around the trimer poles gives
\begin{align}
    t_{0}^h(E,k,p) = \sum_{i}{\frac{R_i(B_3^{(i)},k,p)}{E + B_3^{(i)}}} + \cdots,
\label{eq:t0h-nearpole}
\end{align}
where $i=1,2,\cdots$ labels the trimer poles. $R_i(B_3^{(i)},k,p)$ and $-B_3^{(i)}$ are the residues and energy locations of each pole, respectively.  $R_i(B_3^{(i)},k,p)$ needs to be determined in order to include trimer pole(s) in four-body calculations through the Cauchy principal value prescription, whose details are given in Appendix~\ref{app:Cauchy}. To find the expressions of $h$ and $R_i(B_3^{(i)},k,p)$, one can split $t^h_{\ell}(E, k, p)$ into two terms~\cite{Vanasse_2017}, which are shown in Fig.~\ref{fig:dimer-boson-rewritten} and read:
\begin{align}
    t_{0}^h(E,k,p) = t^{h = 0}_{0}(E,k, p) -\mathcal{G}(E, k)D_{t}(E)\mathcal{G}(E, p),
    \label{eqn:dimer-boson rewritten}
\end{align}
where the first term on the right-hand side, $t^{h = 0}_{0}(E, k, p)$, does not contain any three-boson contact interaction, and the second term contains the non-perturbative contribution of the three-boson contact interaction.
$\mathcal{G}(E, p)$ is the three-boson vertex function defined through the integral equation 
\begin{align}
    \mathcal{G}(E, p) = \omega + \int_{0}^{
\Lambda} \frac{q^2dq}{2\pi^2} \mathcal{G}(E, q)K_{0}^{h=0}(E,q,p),
\label{eqn:vertex function}
\end{align}
where the superscript $h=0$ indicates that the three-boson force is not included. The diagrammatic representation of Eq.~\eqref{eqn:vertex function} is shown in Fig. \ref{fig:trimer vertex function}. 
The dressed trimer propagator $iD_{t}(E)$ in Eq.~\eqref{eqn:dimer-boson rewritten} is given by the geometric sum in Fig.~\ref{fig:dressed trimer} and reads
\begin{align}
    iD_{t}(E) &= \frac{i}{\Omega} + \frac{i}{\Omega}(-\frac{\omega^2}{\Omega}\Sigma_{\Lambda}(E)) + \frac{i}{\Omega}(-\frac{\omega^2}{\Omega}\Sigma_{\Lambda}(E))^2 + \cdots \nonumber \\
    &= \frac{i}{\Omega}\frac{1}{1 - h\Sigma_{\Lambda}(E)},
    \label{eq:dressed-trimer-prop}
\end{align}
where Eq.~\eqref{eqn:hmatching} is used. $\Sigma_{\Lambda}(E)$ is defined as
\begin{align}
    \Sigma_{\Lambda}(E) = -\int_{0}^{\Lambda}\frac{q^2dq}{2\pi^2}\mathcal{G}(E,q)D_d(E - \frac{q^2}{2m}, q).
    \label{eq:sigmaE}
\end{align}
In order to make $t^{h}_{0}(E, k, p)$ cutoff-independent, one needs to fit $h$ at each $\Lambda$ to the trimer poles of $t^{h}_{0}(E, k, p)$ at $E=-B_3^{(i)}$ . Because $t^{h=0}_{0}(E, k, p)$ does not have poles at $E=-B_3^{(i)}$ except for special cutoffs, the dressed trimer propagator, $iD_t{(E)}$, and therefore also the second term on the right-hand side of Eq.~\eqref{eqn:dimer-boson rewritten} are singular at $E=-B_3^{(i)}$. Fixing the location of a trimer pole in $iD_t(E)$ at $E = -B_3^{(i)}$ yields the expression for $h$:
\begin{align}
    h = \frac{1}{\Sigma_{\Lambda}(-B_3^{(i)})}.
    \label{eq:hfit}
\end{align}
Using this three-boson force in $iD_t(E)$ and plugging $iD_t(E)$ into Eq.~\eqref{eqn:dimer-boson rewritten} yields the expression for $R_i(B_3^{(i)},k,p)$:
\begin{align}
    R_i(B_3^{(i)},k,p) = -\frac{\mathcal{G}(-B_3^{(i)}, k)\mathcal{G}(-B_3^{(i)}, p)}{\omega^2\Sigma'_{\Lambda}(-B_3^{(i)})},
    \label{eqn:trimer residues}
\end{align}
which is separable in $k$ and $p$. See Ref.~\cite{Ji_2012} for a different derivation of the separable structure of the three-body residue.

As $\Lambda$ goes to infinity, $h$ shows a limit-cycle behavior~\cite{Braaten_2006} whose phase is fixed by $-B_3^{(i)}$. At each $\Lambda$, the locations of all trimer poles are fixed once $h$ is determined. At some special values of $\Lambda$, $\abs{\Sigma_{\Lambda}(-B_3^{(i)})} \to \infty$ and $\abs{h} \to 0$. The trimer poles of $t^{h=0}_{0}(E, k, p)$ at those cutoffs occur at $E=-B_3^{(i)}$ and Eq.~\eqref{eqn:trimer residues} may be ill-defined. Those special cutoffs are avoided in these calculations to prevent numerical instabilities.  It is also noteworthy that the vertex function $\mathcal{G}(-B_3^{(i)}, p)$ satisfies
\begin{align}
    \mathcal{G}(-B_3^{(i)}, p) = \int_{0}^{\Lambda} \frac{q^2dq}{2\pi^2} \mathcal{G}(-B_3^{(i)}, q)K_{0}^h(-B_3^{(i)},q,p),
    \label{eq:G_eigenv}
\end{align}
where $K_{0}^h(-B_3^{(i)},q,p)$  includes the three-boson force. Eq.~\eqref{eq:G_eigenv} can be checked by plugging Eq.~\eqref{eq:K_l^h} into Eq.~\eqref{eq:G_eigenv} and using Eqs~\eqref{eq:sigmaE} and~\eqref{eq:hfit} to simplify the expressions.
This means $\mathcal{G}(-B_3^{(i)}, p)$ i is an eigenvector, with an eigenvalue equal to one, of the kernel of Eq.~\eqref{eqn:integral equation for three-boson scatterings} in S wave.

\section{Four-Boson Systems}
\label{sec:4bosonSys}

\subsection{Diagrammatic and operator form of the four-boson integral equation}
\label{sec:4bodySys-formal} 
\begin{figure}[htb!]
    \centering    
    \includegraphics[trim={10cm 18cm 10cm 10cm},clip,width=7in]{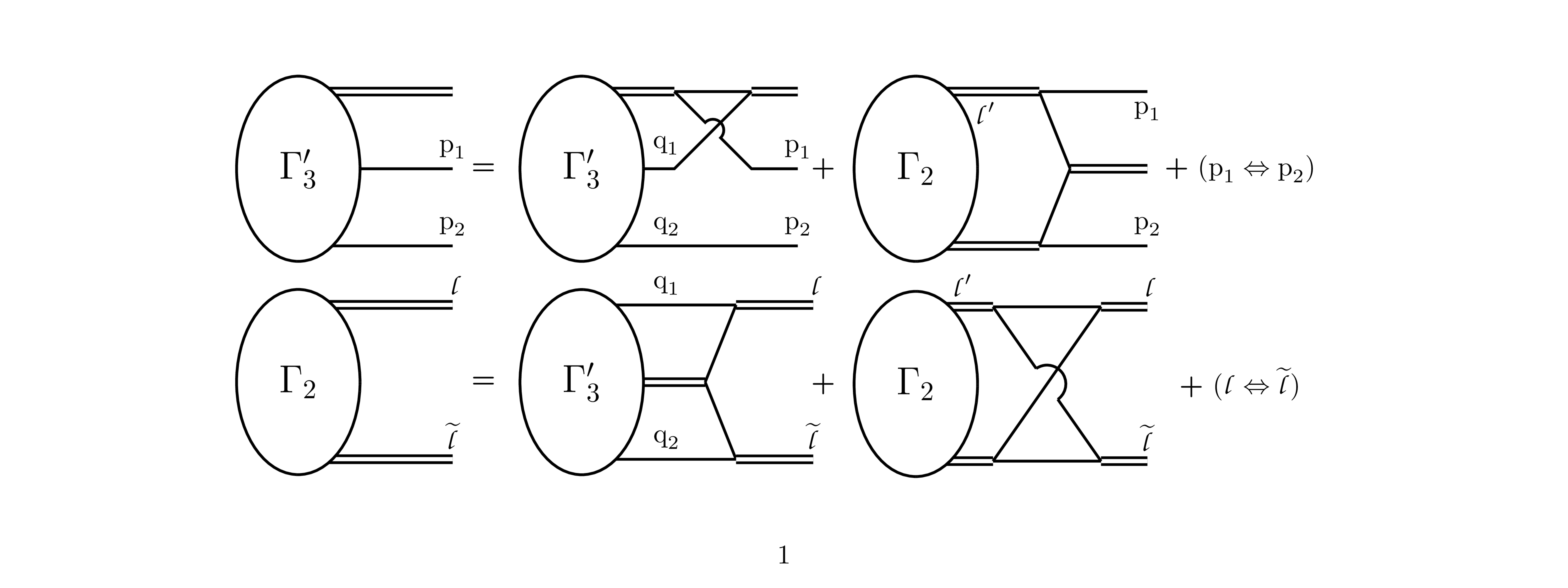}
    \caption{Homogenous part of the four-boson integral equation without a three-boson force. $\Gamma'_3$ and $\Gamma_2$ are the four-boson amplitudes with outgoing dimer-two-single-boson and dimer-dimer states, respectively. Four momenta are indicated.}
    \label{fig:Intg_K3}
\end{figure}

The homogeneous four-boson integral equation~\cite{PhysRevA.73.032724} without three-boson forces is shown diagrammatically in Fig.~\ref{fig:Intg_K3}.  Inhomogeneous terms are not needed for studying tetramer binding energies, but can be added for scattering. Four-momenta are indicated in Fig.~\ref{fig:Intg_K3} and will be explained below. (See Appendix~\ref{app:baddiagrams} for a simpler set of four-boson diagrams and why they are not used to form a four-body integral equation.) 
The operator form of this four-boson integral equation reads
\begin{equation}
    \begin{aligned}
    \Gamma'_3& =(1 + \mathcal{P}_3) \left(K_{33} \Gamma'_3 + K_{32} \Gamma_2\right)\\
     \Gamma_2& = (1 + \mathcal{P}_2) \left(K_{23} \Gamma'_3 + K_{22} \Gamma_2\right),
     \label{eq:formal_Intg_K3}
\end{aligned}
\end{equation}
where $\Gamma'_3$ and $\Gamma_2$ are the four-boson amplitudes with outgoing dimer-two-single-boson and dimer-dimer states, respectively. $\mathcal{P}_3$ ($\mathcal{P}_2$) is the permutation operator that interchanges the two outgoing single bosons (dimers). Similar to the three-boson case, one can invoke the residue theorem on loop integrals involving $\Gamma'_3$ and pick up the single-boson poles, whose four-momenta are $\mathrm{p}_n$ or $\mathrm{q}_n$, $n=$1,~2, as indicated in Fig.~\ref{fig:Intg_K3}. This leads to a closed system of equations 
corresponding to Eq.~\eqref{eq:formal_Intg_K3} with single-boson momenta $\textrm{p}_n$ and $\textrm{q}_n$ being on-shell, i.e., $\textrm{p}_n = \{p_n^2/(2m), \vect{p}_n\}$ and $\textrm{q}_n = \{q_n^2/(2m), \vect{q}_n\}$. For loops involving $\Gamma_2$, a branch cut is associated with each of the two dimer propagators, one in the upper half and the other in the lower half of the energy complex plane. It is thus difficult to use the residue theorem on energy integrals that involve $\Gamma_2$. 
Therefore, regarding the four momenta  $\text{\lfont l }= \{E/2 + l_0, ~\vect{l}\}$ and $\widetilde{\text{\lfont l }} = \{E/2 - l_0, -\vect{l}\}$, where $E$ is now the total energy of the four-boson system, both $l_0$ and $\vect{l}$ are taken as integration variables.
To avoid the singularities along the real axis of $l_0$, one can deform the contour for $l_0$ and integrate $l_0$ from $-i\infty$ to $i\infty$ instead. See Ref.~\cite{PhysRevA.73.032724} about this deformation. Summarizing the above, at total energy $E$ in the four-boson CM frame, $\Gamma'_3$ depends on $\vect{p}_1$ and $~\vect{p}_2$, and $\Gamma_2$ depends on $l_0$ and $\vect{l}$. These momenta will be referred to as \fourBodyMom.
A sharp cutoff is used for these momenta in this paper and no interpolation is needed when discretizing them.
\begin{figure}[htb]
    \centering    
    \includegraphics[trim={10cm 14cm 80cm 10cm},clip,width=5in]{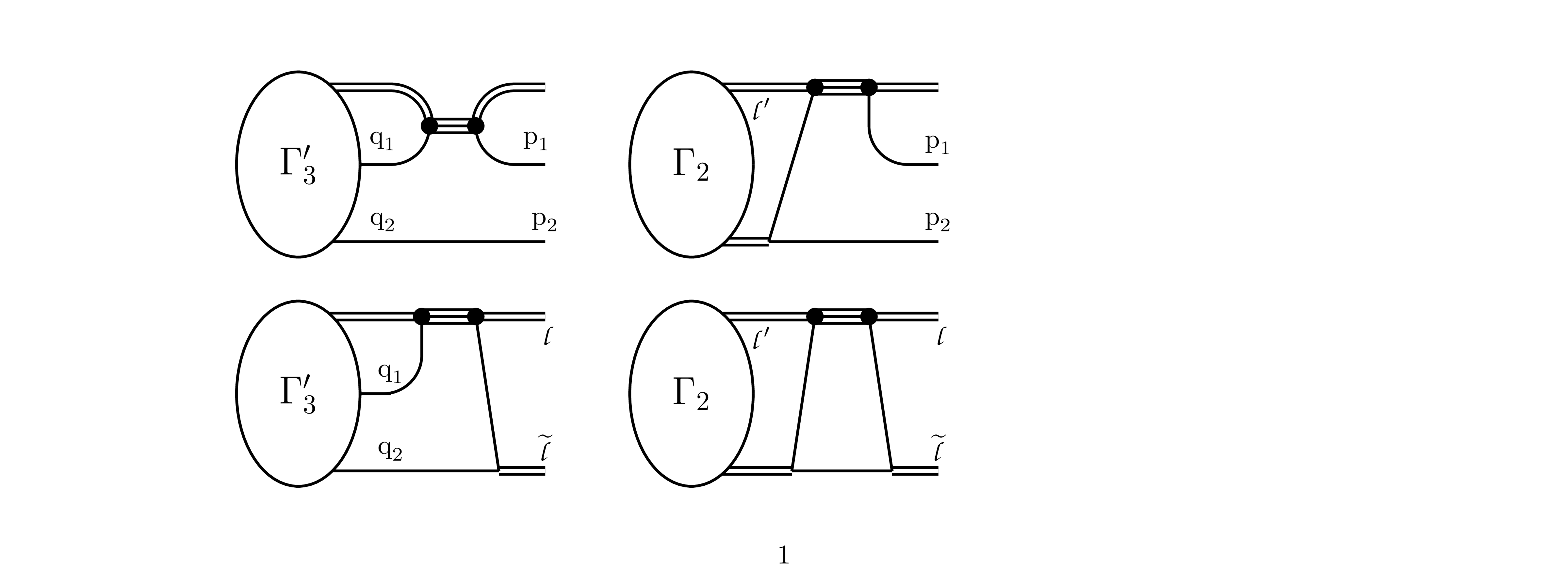}
    \caption{Four-boson Feynman diagrams with a three-boson force. These diagrams should be included on the right-hand side of the integral equation shown in Fig.~\ref{fig:Intg_K3} when a three-boson force is needed.}
    \label{fig:4body_withh}
\end{figure}

Four-boson diagrams with a three-boson contact interaction, shown in Fig.~\ref{fig:4body_withh}, should be added to the kernel of the integral equation shown in Fig.~\ref{fig:Intg_K3}. However, with \fourBodyMom and their sharp cutoff $\Lambda'$, it is tricky to determine the strength of the three-boson force, $h$, to be used in a four-body calculation because $h$ depends on the sharp cutoff $\Lambda$ on the relative momentum between the dimer and single boson, as opposed to the cutoff on \fourBodyMom. (This was not a problem in Ref.~\cite{PhysRevA.73.032724} as no three-body contact interaction was used.) This seems to motivate using Jacobi momenta instead of \fourBodyMom, similar to how Ref.~\cite{Platter:2004he} solved the FY equation (also see Ref.~\cite{KAMADA1992205} and~\cite{Gloeckle:1993vr}), though the four-boson integral equation is not closed under the discretized Jacobi momenta, making interpolations unavoidable~\cite{KAMADA1992205}.

In order to address trimer poles in four-boson calculation and avoid interpolations, this paper adopts an approach where both Jacobi momenta and \fourBodyMom are used. 
Inspection of the kernel of the four-boson integral equation reveals that the iterations over only $K_{33}$ consist of an interacting three-boson subsystem and a spectator boson, i.e., $K_{33}$ leaves the (3+1) fragmentation invariant. This means that one can first compute the dimer-single-boson amplitude with Jacobi momentum and then use this amplitude as a subdiagram in the four-boson integral equation, where the \fourBodyMom are used as integration variables. In other words, the \fourBodyMom are used to close the system of linear equations upon discretization.
\begin{figure}[htb]
    \centering    
    \includegraphics[trim={20cm 18cm 10cm 10cm},clip,width=6.3in]{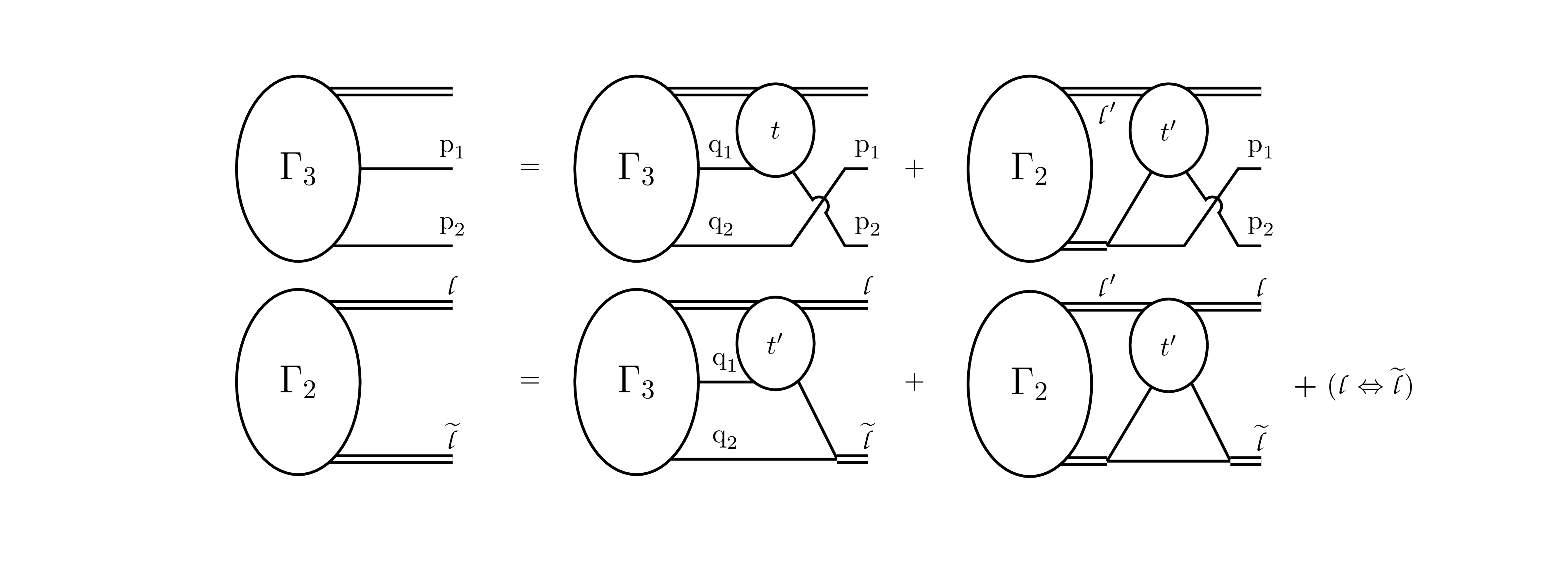}
    \caption{Feynman diagrams for the modified four-boson integral equation in terms of dimer-single-boson scattering amplitudes, which are labeled $t$ if both single-bosons attached are on-shell, or labeled $t'$ if at least one boson attached is off-shell. The definition of the kernel is given by Eqs.~\eqref{eq:formal_Intg_T3} and~\eqref{eq:K's}. }
    \label{fig:Intg_A3}
\end{figure}
The modified four-boson integral equation is shown in Fig.~\ref{fig:Intg_A3}, where the \fourBodyMom are indicated, and the operator form of this integral equation reads
\begin{equation}
    \begin{aligned}
    \Gamma_3& = \left(K'_{33} \Gamma_3 + K'_{32}\Gamma_2\right)\\
     \Gamma_2& = (1 + \mathcal{P}_2) \left(K'_{23} \Gamma_3 + K'_{22} \Gamma_2\right),
     \label{eq:formal_Intg_T3}
\end{aligned}
\end{equation}
where $\Gamma_3$ is related to $\Gamma'_3$ through a permutation $\Gamma'_3 = (1+\mathcal{P}_3)\Gamma_3$ and the kernels are given by
\begin{equation}
    \begin{aligned}
    K'_{23} &\equiv K_{23}\left(\mathrm{1} + T_{33}\right)  \\
    K'_{32} &\equiv \mathcal{P}_3\left(\mathrm{1} + T_{33}\right)K_{32}\\
    K'_{22} &\equiv K_{23}\left(\mathrm{1} + T_{33}\right)K_{32}+ K_{22}\\
    K'_{33} &\equiv \mathcal{P}_3T_{33},
\end{aligned}
\label{eq:K's}
\end{equation}
with 
\begin{align}
    \centering
    T_{33} \equiv K_{33}(1-K_{33})^{-1}.
    \label{eq:T_33}
\end{align}
An algebraic derivation of Eq.~\eqref{eq:formal_Intg_T3} from Eq.~\eqref{eq:formal_Intg_K3} is given in Appendix~\ref{app:four-body}. $K_{23}$ and $K_{32}$ are not further combined with $T_{33}$ because the single boson propagators in $K_{23}$ and $K_{32}$ are generally off-shell. This is also one reason for labeling as $t$ the oval in the top-left component of the kernel in Fig.~\ref{fig:Intg_A3} while labeling as $t'$ the ovals in the other three components. In order to solve Eq.\eqref{eq:formal_Intg_T3}, one can first solve 
\begin{align}
    \Gamma_2 
      = (1 + \mathcal{P}_2)\left[ K'_{23}\left(1 -  K'_{33}\right)^{-1}K'_{32} + K'_{22}\right]\Gamma_2, 
     \label{eq:gamma2_main}
\end{align}
as long as $\left(1 -  K'_{33}\right)$ is invertible at energies under consideration. An algebraic derivation of Eq~\eqref{eq:gamma2_main} from Eq.~\eqref{eq:formal_Intg_K3} can be found in Appendix~\ref{app:four-body}. Eq.~\eqref{eq:gamma2_main} is numerically cheaper to solve for the tetramer binding energies, which are the only observables of interest here. 

As will be shown in the next subsection, the three-boson amplitude $t^{h}_{\ell}(E,k,p)$ appears in the four-boson integral equation explicitly through the matrix element of $T_{33}$ defined in Eq.~\eqref{eq:T_33}. This echoes Weinberg's idea of expressing four-body amplitude in terms of two- and three-body amplitudes~\cite{PhysRev.133.B232}. In addition, since the residues of $t^{h}_{\ell}(E,k,p)$ at the trimer poles are already given by Eq.~\eqref{eqn:trimer residues},  these trimer poles can be included or subtracted in Eqs.~\eqref{eq:formal_Intg_T3} and \eqref{eq:gamma2_main} using the Cauchy principal value prescription detailed in Appendix~\ref{app:Cauchy}. In particular, tetramers that are bound states at low cutoffs become resonances at large cutoffs if the deep trimer poles are included, whereas these tetramers remain bound at large cutoffs if the deep trimer poles are subtracted. Compared to previous studies limited to relatively low cutoffs where no deep trimer exists, this method makes it possible to solve for tetramer binding energies (and decay widths if the tetramers are considered resonances) at higher cutoffs. Scatterings can also be studied using this method.

\subsection{Integral equation under a partial-wave basis}
\label{sec:4bodySys-expressions}
 This section shows the derivation of the integral form of the four-boson integral equation in a partial-wave basis.\footnote{Some of the expressions in this section have been given by Ref.~\cite{PhysRevA.73.032724}.} In principle, the four-body integral equation could also be solved using explicit angular variables, where the analytic form of the four-body kernel is more straightforward. However, numerically it is much more efficient to diagonalize the kernel matrix in a partial wave basis by including only the lowest few partial waves that dominate at low energies. As will be shown later, reasonably good results for four-boson binding energies can be obtained using only the lowest three partial waves, where the diagonalization is about $30$ times faster than using, for example, 10 points\footnote{The complexity of diagonalizing an $n$ by $n$ matrix is $O(n^3)$ in general.} for each angular variable. Such speedup is far from negligible in a four-body calculation. 
 
The transformation between Jacobi momenta, $\vect{p}^J_1$ and $\vect{p}_2^J$, and the \fourBodyMom, $\vect{p}_1$ and $\vect{p}_2$, are given by (similar for $\vect{q}^J_1$ and $\vect{q}_2^J$)
 \begin{align}
        \vect{p}^J_1(\vect{p}_1, \vect{p}_2) &= \vect{p}_1 + \frac{1}{3}\vect{p}_2, \qquad \vect{p}^J_2(\vect{p}_2) = \vect{p}_2, 
    \label{eq:Jacobi momentum}
\end{align}
where the superscript $J$ refers to Jacobi momenta. $\vect{p}^J_1$ represents the relative momentum between the dimer with a momentum $\vect{p}_1 + \vect{p}_2$ and the single boson with a momentum $\vect{p}_1$. This dimer and single boson compose a three-boson subsystem, and the relative momentum between this three-boson subsystem and the remaining (spectator) boson with a momentum $\vect{p}_2$ is given by $\vect{p}_2^J$. The four-boson state, which consists of one off-shell dimer and two on-shell single bosons, is spanned by the basis $|\vect{k}_1 \vect{k}_2\rangle$, where $\vect{k}_1$ and $ \vect{k}_2$ are either Jacobi or \fourBodyMom in the four-boson CM frame (the dependence on the total energy $E$ of the four-boson system is suppressed).  The magnitude and unit direction of $\vect{p}^J_1$ and $\vect{p}^J_2$ are denoted by
\begin{align}
    p_1^J(\vect{p}_1, \vect{p}_2) &= \abs{\vect{p}_1 + \frac{1}{3}\vect{p}_2}\nonumber\\
    \hatvect{p}_1^J(\vect{p}_1, \vect{p}_2)& = \widehat{\left(\vect{p}_1+\frac{1}{3}\vect{p}_2\right)},
\end{align}
and
\begin{align}
    p_2^J( \vect{p}_2) &= \abs{\vect{p}_2}\nonumber\\
    \hatvect{p}_2^J(\vect{p}_2)& = \hatvect{p}_2.
\end{align}

The dimer-two-single-boson state in the four-body partial-wave basis is
\begin{equation}
    |Lm_L(\ell\lambda)k_1k_2\rangle=\sum_{m,\rho}\CG{\ell}{\lambda}{L}{m}{\rho}{m_L}\int\!\!\frac{d\Omega_{k_1}}{4\pi}\int\!\!\frac{d\Omega_{k_2}}{4\pi}Y_\ell^m(\widehat{\boldsymbol{k}}_1)Y_\lambda^\rho(\widehat{\boldsymbol{k}}_2)|\vect{k}_1\vect{k}_2\rangle,
    \label{eq:pwb-full}
\end{equation}
where $\ell[\lambda]$ and $m[\rho]$ are the angular momentum and its $z$-component associated with $\vect{k}_1[\vect{k}_2]$, respectively. $L$ and $m_L$ are the total angular momentum and its $z$-component, respectively. $\CG{\ell}{\lambda}{L}{m}{\rho}{m_L}$ is a Clebsch–Gordan coefficient, and $Y_\ell^{m}$ are spherical harmonics. The rest of the paper focuses on the four-boson S wave, i.e., $L = m_L = 0$, where the bases simplify into (see Appendix~\ref{app: identities})
\begin{align}
    |(\ell\lambda)k_1k_2\rangle  
    &\equiv|L = 0, m_L = 0, (\ell\lambda)k_1k_2\rangle \nonumber\\ &=\int\!\!\frac{d\Omega_{k_1}}{4\pi}\int\!\!\frac{d\Omega_{k_2}}{4\pi}\delta_{\ell, \lambda}(-1)^{\ell }\sqrt{2\ell + 1}P_\ell\left(\widehat{\boldsymbol{k}}_1\cdot \widehat{\boldsymbol{k}}_2\right)|\vect{k}_1\vect{k}_2\rangle,
    \label{eq:partial wave basis}
\end{align}
where $P_\ell$ is a Legendre polynomial. The dimer-dimer state in the partial wave basis is
\begin{align}
    |Lm_L,E/2 + l_0, l \rangle = \int \frac{d\Omega_{l}}{4\pi}Y_L^{m_L}(\,\hatvect{l}\,)|E/2 + l_0,\vect{l}\rangle.
\end{align}
For four-body S wave, this simplifies into
\begin{align}
    |E/2 + l_0, l \rangle &\equiv |L = 0, m_L = 0,E/2 + l_0, l \rangle \nonumber\\
    &= \int \frac{d\Omega_{l}}{4\pi}|E/2 + l_0,\vect{l}\rangle.
    \label{eq:partial wave basis-l}
\end{align}
Note that $\mathcal{P}_2|E/2 + l_0, l \rangle = |E/2 - l_0, l \rangle$.

One is now ready to compute the matrix elements of the operators defined in Eq.~\eqref{eq:K's}. First, the matrix element of $K_{33}$ in \JacobiPW is given by:
\begin{align}
    K_{33,~\textrm{JJ}}^{(\ell\lambda; \ell'\lambda')}\left(p^J_1, p^J_2; q^J_1,q^J_2\right) &\equiv
    \langle (\ell\lambda) p^J_1 p^J_2 |K_{33}| (\ell'\lambda') q^J_1 q^J_2\rangle \nonumber \\ 
    &= \delta_{\ell, \lambda}\delta_{\ell', \lambda'}\delta_{\ell, \ell'} \frac{2\pi^2}{(q^J_2)^2}\delta\left(q^J_2 - p^J_2\right)K^h_{\ell}\left(E - \frac{(q_2^J)^2}{6m}, p^J_1, q^J_1\right), 
    \label{eq:k33_J}
\end{align}
where $K_{\ell}^h$ is the three-body kernel given by Eq.~\eqref{eq:K_l^h}. The subscript $\textrm{JJ}$ indicates the Jacobi momentum basis. The dependence on the total energy $E$ is suppressed without ambiguity to simplify notations.
The matrix element of $T_{33}$ in \JacobiPW is given by
\begin{align}
     &\quad T_{33,~\textrm{JJ}}^{(\ell\lambda; \ell'\lambda')}\left(p_1^J, p_2^J; q_1^J,q_2^J\right)\nonumber\\
    &\equiv
    \langle (\ell\lambda) p_1^J p_2^J |K_{33}(1-K_{33})^{-1}| (\ell'\lambda') q_1^J q_2^J\rangle \nonumber \\[0.5em]
    &= -\delta_{\ell, \lambda}\delta_{\ell', \lambda'}\delta_{\ell, \ell'}\frac{2\pi^2}{(q^J_2)^2}\delta\left(q^J_2 - p^J_2\right) D_d\left(E - \frac{(q_2^J)^2}{6m}- \frac{(q^J_1)^2}{2m}, \vect{q}^J_1\right)t^h_{\ell}\left(E - \frac{(q_2^J)^2}{6m}, p^J_1, q^J_1\right)\nonumber\\
    &\equiv \delta_{\ell, \lambda}\delta_{\ell', \lambda'}\delta_{\ell, \ell'} \frac{2\pi^2}{(q^J_2)^2}\delta\left(q^J_2 - p^J_2\right) \widetilde{T}_{33,~\textrm{JJ}}^{\ell}\left(p_1^J; q_1^J,q_2^J\right),
    \label{eq:T33_Jacobi}
\end{align}
where the definition of $T_{33}$ given by Eq.~\eqref{eq:T_33} is used on the second line.  On the third line, resolution of the identity in Jacobi momenta is used between $K_{33}$ and $(1-K_{33})^{-1}$ and the resulting expressions are simplified using Eqs.~\eqref{eq:B_l^h},~\eqref{eq:K_l^h}, and the solution to Eq.~\eqref{eqn:integral equation for three-boson scatterings}. On the last line of Eq.~\eqref{eq:T33_Jacobi}, $\widetilde{T}_{33,~\textrm{JJ}}^{\ell}\left(p_1^J; q_1^J,q_2^J\right)$ is defined to simplify expressions later. The matrix elements of $T_{33}$ in \NormPW~are given in Appendix~\ref{app: identities} and will be used below.
Note that at $E - \frac{(q_2^J)^2}{6m} = -B_3^{(i)}$, $t^h_{0}\left(E - \frac{(q_2^J)^2}{6m}, p^J_1, q^J_1\right)$ is singular and its residues are given by Eq.~\eqref{eqn:trimer residues}.
The matrix element of $K'_{33}$ in \NormPW~ reads
\begin{align}
    K^{\prime(\ell\lambda; \ell'\lambda')}_{33,~\textrm{SS}}\left(p_1, p_2; q_1,q_2\right)&\equiv
    \langle (\ell\lambda) p_1 p_2 |\mathcal{P}_3T_{33}| (\ell'\lambda') q_1 q_2\rangle \nonumber \\
    &= T_{33,~\textrm{SS}}^{(\ell\lambda; \ell'\lambda')}\left(p_2, p_1; q_1,q_2\right),
    \label{eq:K33prime_SS}
\end{align}
where the subscript $\textrm{SS}$ indicates the \fourBodyMomdash basis. $T_{33,~\textrm{SS}}^{(\ell\lambda; \ell'\lambda')}$ is related to $T_{33,~\textrm{JJ}}^{(\ell\lambda; \ell'\lambda')}$ through a change of basis, and the expression for $T_{33,~\textrm{SS}}^{(\ell\lambda; \ell'\lambda')}\left(p_2, p_1; q_1,q_2\right)$ is given by Eq.~\eqref{eq:T_33_SS}.
The matrix elements of $K_{32}$, $K_{23}$, and $K_{22}$ in \NormPW~ can be read off the diagrams in Figs.~\ref{fig:Intg_K3} and~\ref{fig:4body_withh}, yielding  
\begin{align}
     &\qquad K_{32,~\textrm{SS}}^{(\ell\lambda)}\left(p_1, p_2; l_0,l\right) \nonumber\\
    &\equiv
    \langle (\ell\lambda) p_1 p_2 | K_{32}|E/2 + l_0, l \rangle \nonumber \\[0.5em]
    &= \delta_{\ell, \lambda}(-1)^\ell\sqrt{2\ell +     1}\iiint \frac{d\Omega_{p_1}}{4\pi}\frac{d\Omega_{p_2}}{4\pi}\frac{d\Omega_{l}}{4\pi}P_\ell\left(\hatvect{p}_1\cdot\hatvect{p}_2\right) \abs{D_d\left(\frac{E}{2}+l_0,\vect{l}\right)}^2\nonumber\\
    & \hspace{1cm}\times\left(-i\sqrt{2}y^3 
    D_N\left(\frac{E}{2}-l_0 - \frac{\vect{p}_2^2}{2m}, \vect{l}+\vect{p}_2\right)
    D_N\left(\frac{E}{2}+l_0 - \frac{\vect{p}_1^2}{2m}, \vect{l}-\vect{p}_1\right)
    \right. \nonumber\\
    & \hspace{1.7cm} \left. + i\sqrt{2}hy
    D_N\left(\frac{E}{2}-l_0 - \frac{\vect{p}_2^2}{2m}, \vect{l}+\vect{p}_2\right)
     \right)  \nonumber \\
    &= \delta_{\ell, \lambda}(-1)^\ell\sqrt{2\ell + 1} \frac{1}{l p_2} Q_{
    \ell
    }\left(\frac{p_2^2 + l^2/2 -m(E/2-l_0)}{l p_2}\right)\abs{D_d\left(\frac{E}{2}+l_0,\vect{l}\right)}^2 \nonumber\\
    &\hspace{1cm}\times
    \left(-i\sqrt{2}y^3\frac{1}{l p_1}Q_{\ell}\left(\frac{p_1^2 + l^2/2 -m(E/2+l_0)}{l p_1}\right)
      + i\sqrt{2}hy\delta_{\ell,0}\right) 
    \label{eq:K_32_normal},
\end{align}
where $D_N\left(p_0, \vect{p}\right)$ is the single-boson propagator
\begin{align}
    D_N\left(p_0, \vect{p}\right) = \frac{1}{p_0 - \frac{p^2}{2m} + i\epsilon}.
\end{align}
In Eq.~\eqref{eq:K_32_normal}, $i\epsilon$ is dropped in the argument of $Q_{\ell}$ because there is no singularity for $l_0$ integrated over the imaginary axis.

The term proportional to $h$ in Eq.~\eqref{eq:K_32_normal} comes from the three-boson counterterm in Fig.~\ref{fig:4body_withh}. Inspection of Eqs.~\eqref{eq:formal_Intg_T3} and~\eqref{eq:gamma2_main} reveals that $\Gamma_2 = \frac{1+\mathcal{P}_2}{2}\Gamma_2$. This suggests the redefinition $K_{32}\to K_{32}(1+\mathcal{P}_2)/2$ and $K_{32,~\textrm{SS}}^{(\ell\lambda)}\left(p_1, p_2; l_0,l\right)$ changes accordingly:
\begin{align}
    K_{32,~\textrm{SS}}^{(\ell\lambda)}\left(p_1, p_2; l_0,l\right) &\rightarrow \frac{1}{2}\left(K_{32,~\textrm{SS}}^{(\ell\lambda)}\left(p_1, p_2; l_0,l\right) + K_{32,~\textrm{SS}}^{(\ell\lambda)}\left(p_1, p_2; -l_0,l\right)\right)\nonumber\\
    & = \Re\left[K_{32,~\textrm{SS}}^{(\ell\lambda)}\left(p_1, p_2; l_0,l\right)\right],~\textrm{if Im(E) = 0}.
\end{align}
This redefinition is convenient (but not necessary) for numerically 
diagonalizing the kernel when the rest of the kernel is purely real. One needs the matrix element of $K_{32}$ in a partial wave basis
\begin{align}
     &\qquad K_{32,~\textrm{JS}}^{(\ell\lambda)}\left(p_1^J, p_2^J; l_0,l\right) \nonumber\\
    &\equiv
    \langle (\ell\lambda) p_1^J p_2^J | K_{32}|E/2 + l_0, l \rangle \nonumber \\[0.5em]
    &= \delta_{\ell, \lambda}(-1)^\ell\sqrt{2\ell + 1}\iiint \frac{d\Omega_{p_1^J}}{4\pi}\frac{d\Omega_{p_2^J}}{4\pi}\frac{d\Omega_{l}}{4\pi}P_\ell\left(\hatvect{p}^J_1\cdot\hatvect{p}^J_2\right) \abs{D_d\left(\frac{E}{2}+l_0,\vect{l}\right)}^2\nonumber\\
    & \hspace{0.5cm}\times\left(
     -i\sqrt{2}y^3 
    D_N\left(\frac{E}{2}-l_0 - \frac{(p_2^J)^2}{2m}, \vect{l}+\vect{p}^J_2\right)D_N\left(\frac{E}{2}+l_0 - \frac{\left(\vect{p}_1^J - \frac{\vect{p}_2^J}{3} \right)^2}{2m}, \vect{l}-\vect{p}_1^J+\frac{\vect{p}_2^J}{3}\right)\right.\nonumber\\
    & \left.\hspace{2cm}+i\sqrt{2}hy
    D_N\left(\frac{E}{2}-l_0 - \frac{(p_2^J)^2}{2m}, \vect{l}+\vect{p}_2^J\right)\vphantom{D_N\left(\frac{E}{2}+l_0 - \frac{\left(\vect{p}_1^J - \frac{1}{3}\vect{p}_2^J\right)^2}{2m}, \vect{l}-\vect{p}_1^J+\frac{1}{3}\vect{p}_2^J\right)}
     \right)  \nonumber \\
    &\begin{aligned}
    &=
    \delta_{\ell, \lambda}(-1)^\ell\sqrt{2\ell + 1} \abs{D_d\left(\frac{E}{2}+l_0,\vect{l}\right)}^2 
    \left( i\sqrt{2}hy\delta_{\ell,0}\frac{1}{l p_2^J} Q_{0}\left(\frac{(p_2^J)^2 + l^2/2 -m(E/2-l_0)}{l p_2^J}\right)\vphantom{Q_{\ell}\left(\frac{(p_1^J)^2 + l^2/2 + (p_2^J)^2/9+\left(\vect{l}\cdot\vect{p}_2^J\right)/3 -m(E/2+l_0)}{p_1^J\abs{\vect{l}+\frac{2}{3}\vect{p}_2^J}}\right)}\right. \\
    &\hspace{1.2cm}\left.-i\sqrt{2}y^3\iint \frac{d\Omega_{p_2^J}}{4\pi}\frac{d\Omega_{l}}{4\pi}P_\ell\left(\widehat{\left(\vect{l}+\frac{2}{3}\vect{p}_2^J\right)}\cdot\hatvect{p}^J_2\right) D_N\left(\frac{E}{2}-l_0 - \frac{(p_2^J)^2}{2m}, \vect{l}+\vect{p}^J_2\right)\right.\\
    &\hspace{1.4cm}\times\left.\frac{1}{p_1^J\abs{\vect{l}+\frac{2}{3}\vect{p}_2^J} }Q_{\ell}\left(\frac{(p_1^J)^2 + l^2/2 + (p_2^J)^2/9+\left(\vect{l}\cdot\vect{p}_2^J\right)/3 -m(E/2+l_0)}{p_1^J\abs{\vect{l}+\frac{2}{3}\vect{p}_2^J}}\right)\right),
     \end{aligned} 
    \label{eq:K_32_Jacobi}
\end{align}
where the subscript $\textrm{JS}$ indicates that the first two arguments (momenta in the bra) are Jacobi momenta while the last two (momenta in the ket) are \fourBodyMom. For the angular integrals on the last line, one can choose $\hatvect{p}_2^J$ along the $z$-axis and numerically integrate over $\Omega_l$.
The matrix element of $T_{33}K_{32}$ in \NormPW is given by:
\begin{align}
     &\qquad (T_{33}K_{32})^{(\ell\lambda)}_{\textrm{SS}}\left(p_1, p_2; l_0,l\right)\nonumber\\
    &\equiv
    \langle (\ell\lambda) p_1 p_2 | T_{33}K_{32}|E/2 + l_0, l \rangle \nonumber \\
    &=\sum_{\ell',\lambda'}\iint_0^{\Lambda} \frac{(k_1^J)^2dk_1^J}{2\pi^2}\frac{(k_2^J)^2dk_2^J}{2\pi^2}
    \langle (\ell\lambda) p_1 p_2 | T_{33}| (\ell'\lambda') k_1^J k_2^J \rangle
    \langle (\ell'\lambda') k_1^J k_2^J |K_{32}|E/2 + l_0, l \rangle \nonumber \\
    &=\delta_{\ell,\lambda}\sum_{\ell'}(-1)^{\ell+\ell'}\sqrt{(2\ell+1)(2\ell'+1)}\int_0^{\Lambda}\frac{(k_1^J)^2dk_1^J}{2\pi^2}\widetilde{T}_{33,~\textrm{SJ}}^{(\ell;\ell')}(p_1,p_2;k_1^J)
    K_{32,~\textrm{JS}}^{(\ell'\ell')}\left(k_1^J, p_2; l_0,l\right),
    \label{eq:T33K32_normal}
\end{align}
where $\iint_{0}^{\Lambda} \equiv \int_{0}^{\Lambda}\int_{0}^{\Lambda}$ and
$\widetilde{T}_{33,~\textrm{SJ}}^{(\ell;\ell')}(p_1,p_2;k_1^J)$ is defined implicitly in Eq.~\eqref{eq:T_33_NJ}. The cutoff for Jacobi momenta (e.g., $k_1^J$ in Eq.~\eqref{eq:T33K32_normal}) is just $\Lambda$ that was used in the dimer-single-boson scattering amplitude.
Eqs.~\eqref{eq:K_32_normal} and~\eqref{eq:T33K32_normal} give the matrix element of $K'_{32}$ in \NormPW:
\begin{align}
    &\qquad K^{\prime(\ell\lambda)}_{32,~\textrm{SS}}\left(p_1, p_2; l_0,l\right)\nonumber\\
    &\equiv
    \langle (\ell\lambda) p_1 p_2 | \mathcal{P}_3(1+T_{33})K_{32}|E/2 + l_0, l \rangle \nonumber \\
    &= K_{32,~\textrm{SS}}^{(\ell\lambda)}\left(p_2, p_1; l_0,l\right) + (T_{33}K_{32})^{(\ell\lambda)}_{\textrm{SS}}\left(p_2, p_1; l_0,l\right).
    \label{eq:K32prime_SS}
\end{align}

The matrix elements of $K_{23}$ and  $K'_{23}$ in \NormPW~ can be obtained in a similar manner:
\begin{align}
     &\qquad K_{23,~\textrm{SS}}^{(\ell\lambda)}\left(l_0,l; q_1, q_2 \right) \nonumber\\
    &\equiv
    \langle E/2 + l_0, l | K_{23}| (\ell\lambda) q_1 q_2  \rangle \nonumber \\[0.5em]
    &= \delta_{\ell, \lambda}(-1)^\ell\sqrt{2\ell + 1}\int_{-1}^{1}\frac{d\left(\hatvect{q}_1\cdot\hatvect{q}_2\right)}{2}P_{\ell}\left(\hatvect{q}_1\cdot\hatvect{q}_2\right)D_d\left(E-\frac{q_1^2}{2m}-\frac{q_2^2}{2m},\vect{q}_1 + \vect{q}_2\right)\nonumber\\
    &\hspace{2cm}\times\sum_{\ell'}(-1)^{\ell'}\sqrt{2\ell' + 1}P_{\ell'}\left(\hatvect{q}_1\cdot\hatvect{q}_2\right)\frac{1}{l p_2} Q_{\ell'}\left(\frac{p_2^2 + l^2/2 -m(E/2-l_0)}{l p_2}\right)\nonumber\\
    &\hspace{2.5cm}\times\left(-i\sqrt{2}y^3\frac{1}{l p_1}Q_{\ell}\left(\frac{p_1^2 + l^2/2 -m(E/2+l_0)}{l p_1}\right)
      + i\sqrt{2}hy\delta_{\ell',0}\right)   
    \label{eq:K_23_normal}
\end{align}
and
\begin{align}
    &\qquad K_{23,~\textrm{SJ}}^{(\ell\lambda)}\left(l_0,l; q^J_1, q^J_2 \right) \nonumber\\
    &\equiv
    \langle E/2 + l_0, l | K_{23}| (\ell\lambda) q^J_1 q^J_2  \rangle \nonumber \\[0.5em]
    &\begin{aligned}
    &=
    \delta_{\ell, \lambda}(-1)^\ell\sqrt{2\ell + 1} D_d\left(E - \frac{(q_2^J)^2}{6m}- \frac{(q^J_1)^2}{2m}, \vect{q}^J_1\right) \\
    &\hspace{0.5cm}\times
    \left( i\sqrt{2}hy\delta_{\ell,0}\frac{1}{l q_2^J} Q_{\ell}\left(\frac{(q_2^J)^2 + l^2/2 -m(E/2-l_0)}{l q_2^J}\right)\right. \\
    &\hspace{1.2cm}\left.-i\sqrt{2}y^3\iint \frac{d\Omega_{q_2^J}}{4\pi}\frac{d\Omega_{l}}{4\pi}P_\ell\left(\widehat{\left(\vect{l}+\frac{2}{3}\vect{q}_2^J\right)}\cdot\hatvect{q}^J_2\right) D_N\left(\frac{E}{2}-l_0 - \frac{(q_2^J)^2}{2m}, \vect{l}+\vect{q}^J_2\right)\right.\\
    &\hspace{1.4cm}\times\left.\frac{1}{q_1^J\abs{\vect{l}+\frac{2}{3}\vect{q}_2^J} }Q_{\ell}\left(\frac{(q_1^J)^2 + l^2/2 + (q_1^J)^2/9+\left(\vect{l}\cdot\vect{q}_2^J\right)/3 -m(E/2+l_0)}{q_1^J\abs{\vect{l}+\frac{2}{3}\vect{q}_2^J}}\right)
      \right)
     \end{aligned} 
    \label{eq:K_23_Jacobi}
\end{align}
One also needs the matrix element of $K_{23}T_{33}$ in \NormPW, which can be computed similarly to that of $T_{33}K_{32}$ in Eq.~\eqref{eq:T33K32_normal}:
\begin{align}
    &\qquad \left(K_{23}T_{33}\right)^{(\ell\lambda)}_{\textrm{SS}}\left(l_0,l; q_1, q_2 \right) \nonumber\\
    &\equiv
    \langle E/2 + l_0, l | K_{23}T_{33}| (\ell\lambda) q_1 q_2  \rangle \nonumber \\[0.5em]
   &=\delta_{\ell,\lambda}\sum_{\ell'}(-1)^{\ell+\ell'}\sqrt{(2\ell+1)(2\ell'+1)}\int_0^{\Lambda}\frac{(k_1^J)^2dk_1^J}{2\pi^2}
    K_{23,~\textrm{SJ}}^{(\ell'\ell')}\left(l_0,l; k_1^J, q_2 \right)
    \widetilde{T}_{33,~\textrm{JS}}^{\ell';\ell}\left( k_1^J;q_1, q_2\right),
    \label{eq:K23T33_normal}
\end{align}
where $\widetilde{T}_{33,~\textrm{JS}}^{\ell';\ell}\left( k_1^J;q_1, q_2\right)$ is defined in Eq.~\eqref{eq:Ttilde_33_JN}.
Using Eq.~\eqref{eq:K_23_normal} and~\eqref{eq:K23T33_normal}, the matrix element of $K'_{23}$ in \NormPW is given by
\begin{align}
    &\qquad K^{\prime(\ell\lambda)}_{23,~\textrm{SS}}\left(l_0,l; q_1, q_2 \right) \nonumber\\
    &\equiv
    \langle E/2 + l_0, l | K_{23}\left(\mathrm{1} + T_{33}\right) | (\ell\lambda) q_1 q_2  \rangle \nonumber \\
    &= K_{23,~\textrm{SS}}^{(\ell\lambda)}\left(l_0,l; q_1, q_2 \right)+\left(K_{23}T_{33}\right)^{(\ell\lambda)}_{\textrm{SS}}\left(l_0,l; q_1, q_2 \right).
    \label{eq:K23prime_SS}
\end{align}

In order to obtain the matrix element of $K'_{22}$, one needs to first compute that of $K_{22}$. Ref.~\cite{PhysRevA.73.032724} calculated the piece without a three-boson force. Including the contribution from the three-boson force yields
\begin{align}
    &\qquad K_{22}(l_0,l;l'_0,l') \nonumber \\
    &\equiv  \langle E/2 + l_0, l |K_{22}|E/2 + l'_0, l' \rangle \nonumber \\
    &= \frac{1}{4}\langle E/2 + l_0, l |(1 + \mathcal{P}_2)K_{22}(1 + \mathcal{P}_2)|E/2 + l'_0, l' \rangle  \nonumber \\
    &=\frac{m^2}{4ll'}\abs{D_d\left(\frac{E}{2}+l_0,\vect{l}\right)}^2\int_0^{\Lambda'}\frac{dk}{2\pi^2}
    \left(\frac{iy^4}{E - \frac{2}{M}\left(k^2 + \frac{l^2}{4} + \frac{l^{\prime2}}{4}\right)} -\frac{ihy^2}{2} \right)\nonumber\\
    & \hspace{0.7cm}\times \ln\left(\frac{\left[(k + l/2)^2 + l^{\prime2}/4 -  E/2\right]^2 - l^{\prime 2}_0}{\left[(k - l/2)^2 + l^{\prime2}/4 - E/2\right]^2 - l^{\prime 2}_0}\right)\ln\left(\frac{\left[(k + l'/2)^2 + l^2/4 -  E/2\right]^2 - l_0^2}{\left[(k - l'/2)^2 + l^2/4 -  E/2\right]^2 - l_0^2}\right),
\label{eq:K22}
\end{align}
where the third line again follows from $\Gamma_2 = \frac{1+\mathcal{P}_2}{2}\Gamma_2$ and a redefinition $K_{22}\to (1 + \mathcal{P}_2)K_{22}(1 + \mathcal{P}_2)$ for convenience. $\Lambda'$ is the sharp cutoff for \fourBodyMom. Finally, using Eqs.~\eqref{eq:T33_Jacobi},~\eqref{eq:K_32_Jacobi},~\eqref{eq:K_23_Jacobi}, and~\eqref{eq:K22} the matrix element of $K'_{22}$ is given by
\begin{align}
&\qquad K'_{22}(l_0,l;l'_0,l') \nonumber \\
&\equiv  \langle E/2 + l_0, l | K_{23}\left(\mathrm{1} + T_{33}\right)K_{32}+ K_{22}|E/2 + l'_0, l' \rangle \nonumber \\
& = K_{22}(l_0,l;l'_0,l') + \sum_{\ell,\lambda}\iint_0^{\Lambda'}\frac{(k_1^J)^2dk_1^J}{2\pi^2}\frac{(k_2^J)^2dk_2^J}{2\pi^2}
K_{23,~\textrm{SJ}}^{(\ell\lambda)}\left(l_0,l; k^J_1, k^J_2 \right) K_{32,~\textrm{JS}}^{(\ell\lambda)}\left(k_1^J, k_2^J; l'_0,l'\right) \nonumber\\
&\hspace{1cm}+\sum_{\ell,\lambda}\iint_0^{\Lambda'}\frac{(k_1^J)^2dk_1^J}{2\pi^2}\frac{(k_2^J)^2dk_2^J}{2\pi^2}\frac{(k^{\prime J}_1)^2dk^{\prime J}_1}{2\pi^2}
\left(K_{23,~\textrm{SJ}}^{(\ell\lambda)}\left(l_0,l; k^J_1, k^J_2 \right)\right.\nonumber \\
&\hspace{6cm}\left.\times
\widetilde{T}_{33,~\textrm{JJ}}^{\ell}\left(k_1^J; k^{\prime J}_1,k_2^J\right)
K_{32,~\textrm{JS}}^{(\ell\lambda)}\left(k^{\prime J}_1, k_2^J; l'_0,l'\right) \right).
\label{eq:K22'}
\end{align}
Using Eqs.~\eqref{eq:K33prime_SS},~\eqref{eq:K32prime_SS},~\eqref{eq:K23prime_SS}, and~\eqref{eq:K22'}, the integral form of Eq.~\eqref{eq:formal_Intg_T3} in \NormPW reads
\begin{align}
    \Gamma_3^{\ell,\lambda}(p_1, p_2) &= \iint_0^{{\Lambda'}}\frac{q_1^2dq_1}{2\pi^2}\frac{q_2^2dq_2}{2\pi^2}K^{\prime(\ell,\lambda;\ell'\lambda')}_{33,~\textrm{SS}}(p_1,p_2;q_1,q_2)\Gamma_3^{\ell',\lambda'}(q_1,q_2)\nonumber\\
    &\qquad+ \int_0^{{\Lambda'}}\frac{l^{\prime2}dl'}{2\pi^2}\int_{-i\Lambda_E}^{i\Lambda_E}\frac{dl'_0}{2\pi}K^{\prime(\ell,\lambda)}_{32,~\textrm{SS}}(p_1,p_2;l'_0,l')\Gamma_2(l'_0,l')\nonumber\\
    \Gamma_2(l_0,l) &= \iint_0^{{\Lambda'}}\frac{q_1^2dq_1}{2\pi^2}\frac{q_2^2dq_2}{2\pi^2}\left(K^{\prime(\ell'\lambda')}_{23,~\textrm{SS}}(l_0,l;q_1,q_2) + (l_0 \leftrightarrow -l_0)\right)\Gamma_3^{\ell',\lambda'}(q_1,q_2)\nonumber\\
    &\qquad+ \int_0^{{\Lambda'}}\frac{l^{\prime2}dl'}{2\pi^2}\int_{-i\Lambda_E}^{i\Lambda_E}\frac{dl'_0}{2\pi}\left(K'_{22}(l_0,l;l'_0,l')+ (l_0 \leftrightarrow -l_0)\right)\Gamma_2(l'_0,l'), 
    \label{eq:4body-intg-full}
\end{align}
where
\begin{align}
    \Gamma_3^{\ell,\lambda}(p_1,p_2) &\equiv \Gamma_3^{\ell,\lambda}(p_1,p_2;\{\boldsymbol{\alpha}\})\nonumber\\ 
    &\equiv \langle(\ell\lambda)p_1p_2|\Gamma_3| \{\boldsymbol{\alpha}\}\rangle\nonumber\\
    \Gamma_2(l_0,l) &\equiv \Gamma_2(l_0,l;\{\boldsymbol{\alpha}\})\nonumber\\
    &\equiv \langle E/2+l_0,l|\Gamma_2| \{\boldsymbol{\alpha}\}\rangle,
\end{align}
with an incoming four-boson state $|\{\boldsymbol{\alpha}\}\rangle$.
In Eq.~\eqref{eq:4body-intg-full}, repeated partial-wave indices are summed over.  A sharp cutoff $i\Lambda_E$ is used for integrating over $l_0$. One can solve the matrix form of Eq.~\eqref{eq:4body-intg-full} directly. Alternatively, one can first decouple the matrix form of Eq.~\eqref{eq:4body-intg-full}, which results in the matrix form of Eq.\eqref{eq:gamma2_main} that can be solved instead. Upon discretization, either equation gives a closed system of linear equations.

\subsection{Implementation of the diagrammatic approach}
\label{sec:4bodySys-implementation}
In order to find the binding energies (and decay widths) of the four-boson bound states (resonances), one can either compute the eigenvalues of the kernel in Eq.~\eqref{eq:4body-intg-full} or its alternative form in Eq.~\eqref{eq:gamma2_main}. A bound state (resonance) is found when an eigenvalue equals one. 
Before computing the eigenvalues of the kernel, one first needs to specify the angular momentum cutoff $\ell_{\textrm{max}}$, the momentum cutoffs for Jacobi and \fourBodyMom $\Lambda$ and $\Lambda'$, respectively, the energy cutoff $\Lambda_E$, and the observables used to fit the two- and three-boson forces.
The calculation in this paper uses the same value for the two momentum cutoffs, i.e., $\Lambda = \Lambda'$, and takes $\Lambda_E = 4\Lambda^2/m$ as the energy cutoff. $\ell_{\textrm{max}} = 2$ is used as the angular momentum cutoff. See Appendix~\ref{app:diffcutoffs} for a more detailed discussion on cutoffs. 

Grisenti et al.~\cite{PhysRevLett.85.2284} measured the $^4$He dimer bond length of $\langle r\rangle=52\pm4$~\r{A}. Using $\langle r\rangle$ and the zero-range approximation, they derived the $^4$He dimer binding energy of $B_2=1.1^{+0.3}_{-0.2}$~mK and the $^4$He atom-atom scattering length of $a_{^4\textrm{He}} =104^{+8}_{-18}$~\r{A}, which is much greater than the typical length scale $l_{^4\textrm{He}}$ of the system set by its effective range $r_e  \approx 7.5$~\r{A}~\cite{Qin_2021,Ji_2013} and van der Waals length $\approx 5.4$~\r{A}. This gives the typical low-energy scale of the EFT that describes cold $^4$He atoms $\hbar^2/(ml_{^4\textrm{He}}^2) \approx 400$~mK~\cite{Platter:2004he, Braaten_2006}, and the LO EFT prediction of an observable is expected to be accurate to $l_{^4\textrm{He}}/a_{^4\textrm{He}} \approx 10\%$. Larger clusters of three and/or four $^4$He atoms were studied experimentally in Ref.~\cite{Bruch2002,Ferlaino2009,Kievsky_2011,Kunitski_2015}.

The $^4$He dimer binding energy has been calculated using the TTY (Tang, Toennies, and Yiu) potential~\cite{TTY} by Roudnev and Yakovlev~\cite{ROUDNEV200097} and Kolganova et al.~\cite{Kolganova_2011}, who both found $B_2\approx 1.31$~mK. The trimer ground and excited state binding energies ($B_3^{(0)}$ and $B_3^{(1)}$, respectively) were calculated by Blume and Greene~\cite{Blume2000MonteCH} using the LM2M2 potential~\cite{LM2M2} by combining Monte Carlo (MC) methods with the adiabatic hyperspherical approximation. Blume and Greene~\cite{Blume2000MonteCH} obtained $B_3^{(1)} = 2.186$~mK and $B_3^{(0)} = 125.5$~mK. Also using the LM2M2 potential, Hiyama and Kamimura~\cite{Hiyama_2012} performed a variational calculation and found $B_3^{(1)} = 2.2706$~mK and $B_3^{(0)} = 126.40$~mK. Very similar results are obtained using the Faddeev equation with the LM2M2 potential (see, e.g., Lazauskas and Carbonell~\cite{Lazauskas_2006}, who also used the FY equation to study four-boson systems).
Using an EFT approach, Qin and Vanasse~\cite{Qin_2021} obtained at LO $B_3^{(1)} = 1.723B_2$ and $B_3^{(0)} = 97.12B_2$ by fitting their two-boson force to $B_2 = 1.312262$~mK~\cite{ROUDNEV200097} and three-boson force to the $^4$He trimer-atom scattering length of $1.205(mB_2)^{-1/2}$, as determined by Roudnev~\cite{ROUDNEV200395} with the TTY potential. 
 
In this paper, the three-boson force $h$ is first fit to $B_3^{(1)} = 1.767~B_2$~\cite{Blume2000MonteCH,Lewerenz1997} with $B_3^{(1)} = 2.186$~mK~\cite{Blume2000MonteCH}, similar to the fits by Platter et al.~\cite{Platter:2004he}. Using this three-boson force, $B_3^{(0)}$ converges to $103.9~B_2$ as $\Lambda \to \infty$.  Different from Platter et al.~\cite{Platter:2004he}, the three-boson force $h$ is then re-fit to $B_3^{(0)} = 103.9~B_2$ at all cutoffs. This removes the cutoff dependence of $B_3^{(0)}$ and its impact on the cutoff dependence of the tetramer binding energies while ensuring that $B_3^{(1)}$ converges to $1.767~B_2$ at large $\Lambda$.
\begin{figure}[htb!]
    \centering
    \includegraphics[width = 5in]{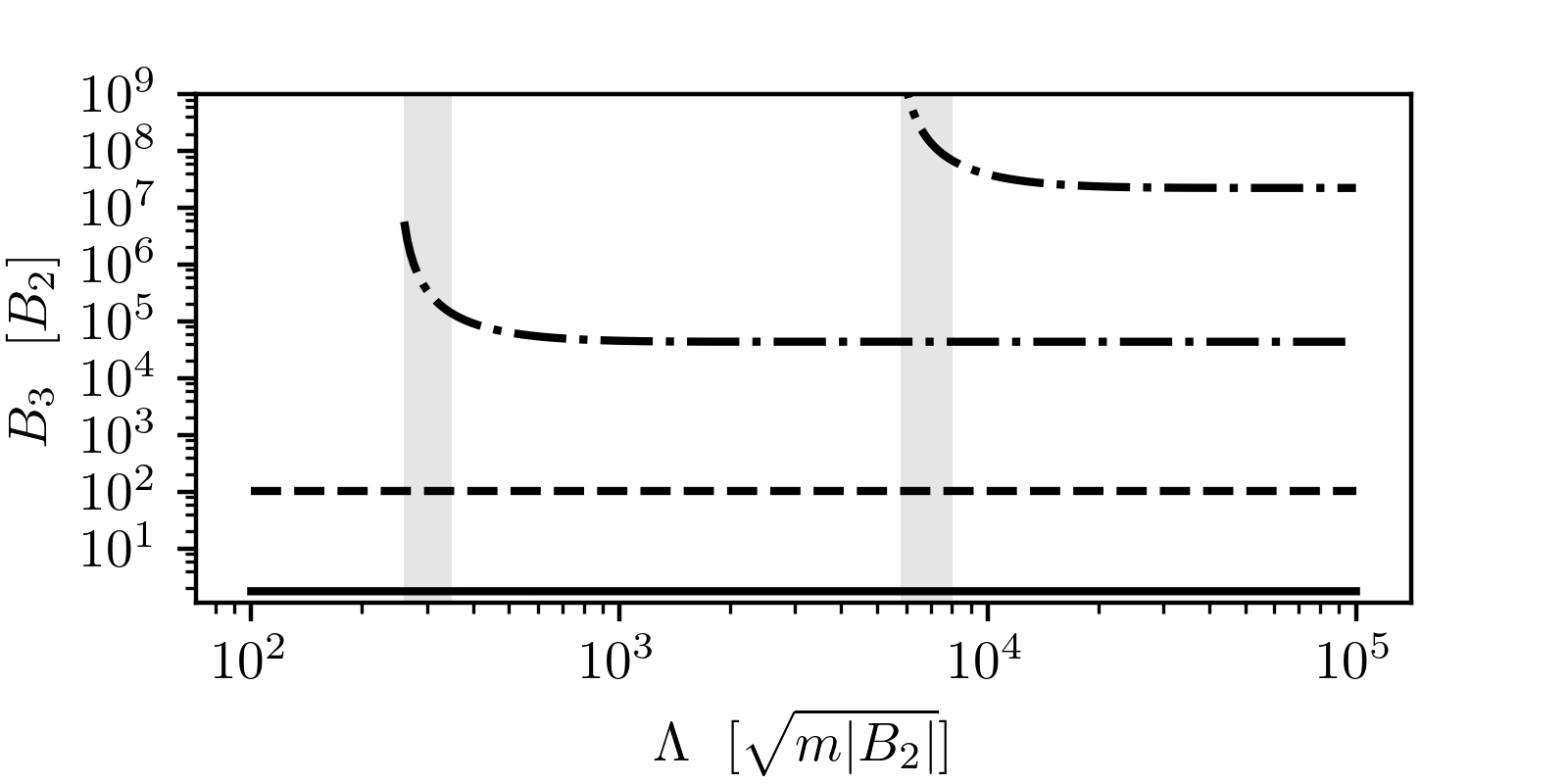}
    \caption{Trimer binding energies as a function of $\Lambda$ for cold $^4$He atoms. The dashed line corresponds to the trimer ground state with $B_3^{(0)} = 103.9B_2$, which is used to fit the three-boson force. The solid line corresponds to the trimer excited state. The dot-dashed lines represent the first two deeply bound trimer states. The gray bands mark the regions where a new trimer state appears with its binding momentum higher than the cutoff, i.e., $\sqrt{mB_3^{(n)}}>\Lambda$, $n=-1,~-2$. }
    \label{fig:B3-all}
\end{figure}
Fig.~\ref{fig:B3-all} shows the cutoff dependence of the trimer binding energies for cold $^4$He atoms. $h$ is fit to the trimer ground state with $B_3^{(0)} = 103.9B_2$, and $B_3^{(1)}$ converges to $1.767B_2$ as $\Lambda \to \infty$. Only these two trimer states exist for $\Lambda \lesssim 250\sqrt{mB_2}\approx\Lambda_t$. As the cutoff increases, more trimer states, referred to as deep (or deeply bound) trimers, appear and their binding energies are at least two orders of magnitude larger than the typical low-energy scale of the EFT of $\hbar^2/(ml_{^4\textrm{He}}^2) \approx 400$~mK. This means that these deep trimers are beyond the cutoff of the EFT, and the trimer with a binding $B_3^{(0)} = 103.9B_2$ is still referred to as the trimer ground state here. The binding energies of the first two deeply bound trimer states are plotted as a function of $\Lambda$ in Fig.~\ref{fig:B3-all} and converge to $B_3^{(-1)} = 4.36\times 10^4 B_2$ and  $B_3^{(-2)} = 2.23\times 10^7B_2$. The bands in Fig.~\ref{fig:B3-all} indicate the appearance of deeply bound trimers and cover the range of $\Lambda$ where $\Lambda$ is smaller than the binding momentum of each deeply bound trimer, i.e.,  where $\Lambda <\sqrt{mB_3^{(n)}}$, $n=-1$ or $-2$. These two deep trimer poles, once they appear in the four-boson calculation at large cutoffs, are included or subtracted using the Cauchy principal value prescription explained in Appendix~\ref{app:Cauchy}. Including them makes the eigenvalues of the four-body kernel complex at $E>-B_3^{(-1)}$ and the tetramers associated with the trimer ground state become resonances. The binding energies and decay widths of these tetramers are found perturbatively by expanding the eigenvalues, explained in Appendix~\ref{app: Eigenvalues}. Their complex binding energies are denoted by $B_4^{(m)}$ with $m=$ 0 or 1, corresponding to the tetramer ground and excited state, respectively, with
\begin{align}
    B_4^{(m)} = E_4^{(m)} + \frac{i\Gamma_4^{(m)}}{2},
\end{align}
where both $E_4^{(m)}$ and $\Gamma_4^{(m)}$ are real and represent the binding energies and decay widths of the tetramers, respectively. One can also choose to subtract the deep trimer poles, which forbids the tetramer bound states of cold $^4$He atoms from becoming resonances at large cutoffs. As shown in Appendix~\ref{app:Cauchy}, this can be done by dropping the $i\pi$ terms when Cauchy principal value prescription is used in the four-boson integral equation.  

The behavior of $E_4^{(m)}$ and $\Gamma_4^{(m)}$ near the unitary limit is studied by changing the ratio $B_2/B_3^{(0)}$. In the unitary limit, $B_2$ goes to zero and there is no three-body scale due to the Efimov effect. The three-boson force $h$ is still fit to a fixed $B_3^{(0)}$ at each cutoff and the tetramers associated with this cutoff-independent trimer are studied. In this case, at least one deep trimer is included and the tetramers associated with $B_3^{(0)}$ are treated as resonances.

\section{Results on Four-boson Systems}
\label{sec:results}
\begin{figure}[htb!]
\centering
\begin{minipage}[c]{\textwidth}
\centering
    \includegraphics[width=12cm]{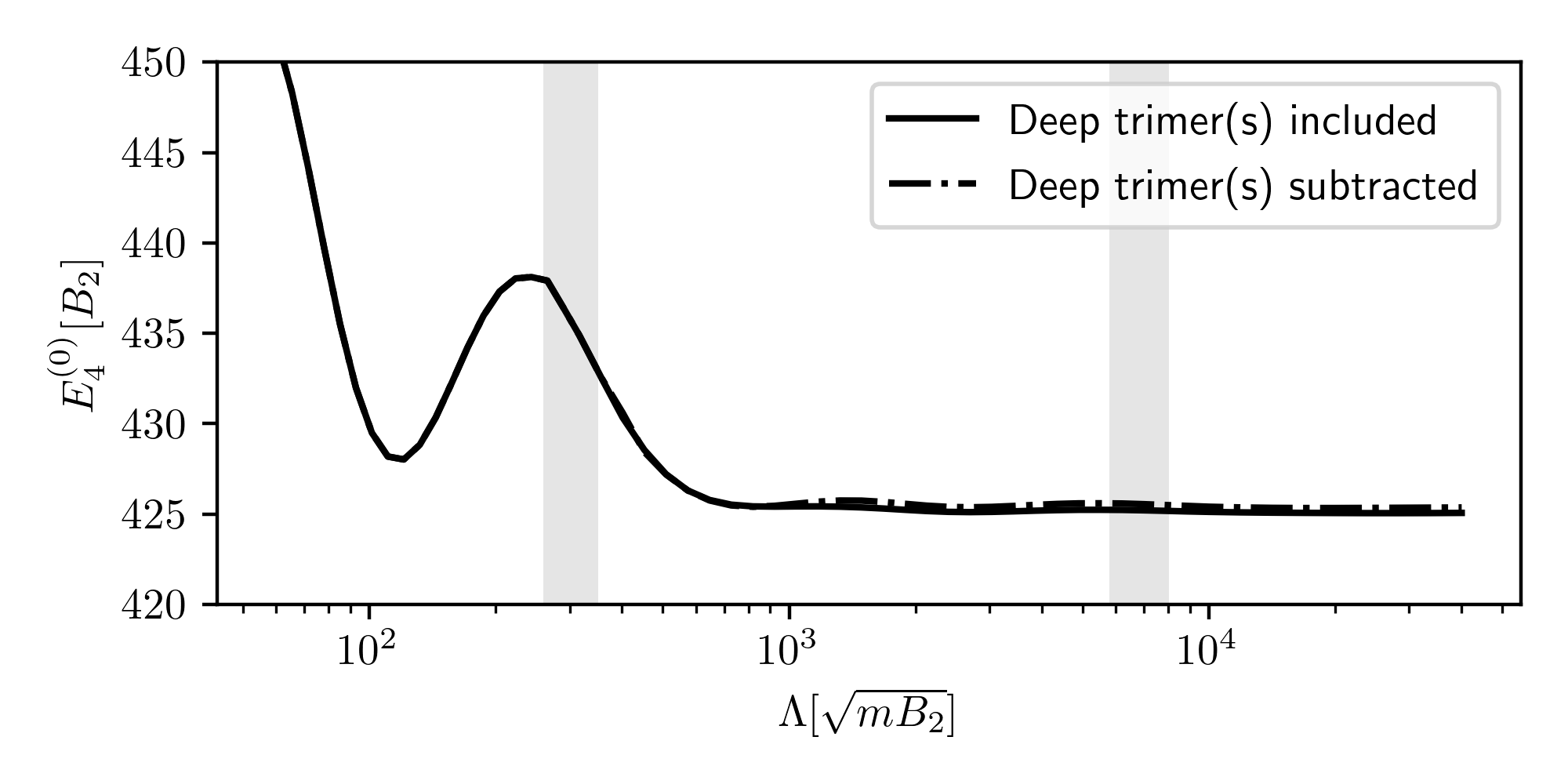}
\end{minipage}
\vfill
\begin{minipage}[c]{\textwidth}
\centering
    \includegraphics[width=12cm,trim={0.3cm 0 0 0},clip]{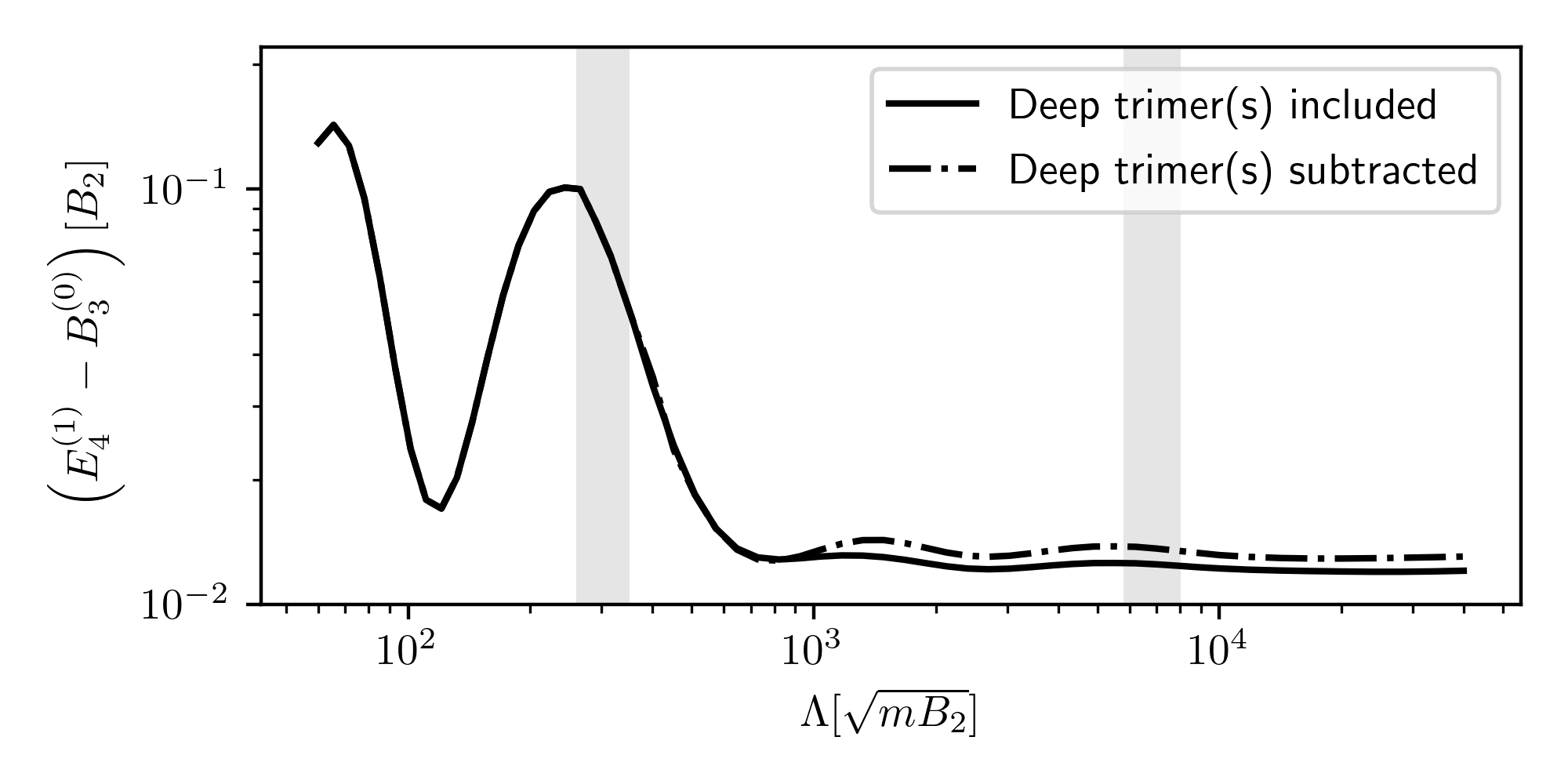}
\end{minipage}
\caption{Cutoff dependence of $E_4^{(0)}$(top), and $(E_4^{(1)} - B_3^{(0)})$ (bottom). The gray bands are the same as of Fig.~\ref{fig:B3-all}. }
\label{fig: 4body-bindings}
\end{figure}
\begin{figure}[htb!]
\centering
\begin{minipage}[c]{\textwidth}
\centering
    \includegraphics[width=10cm]{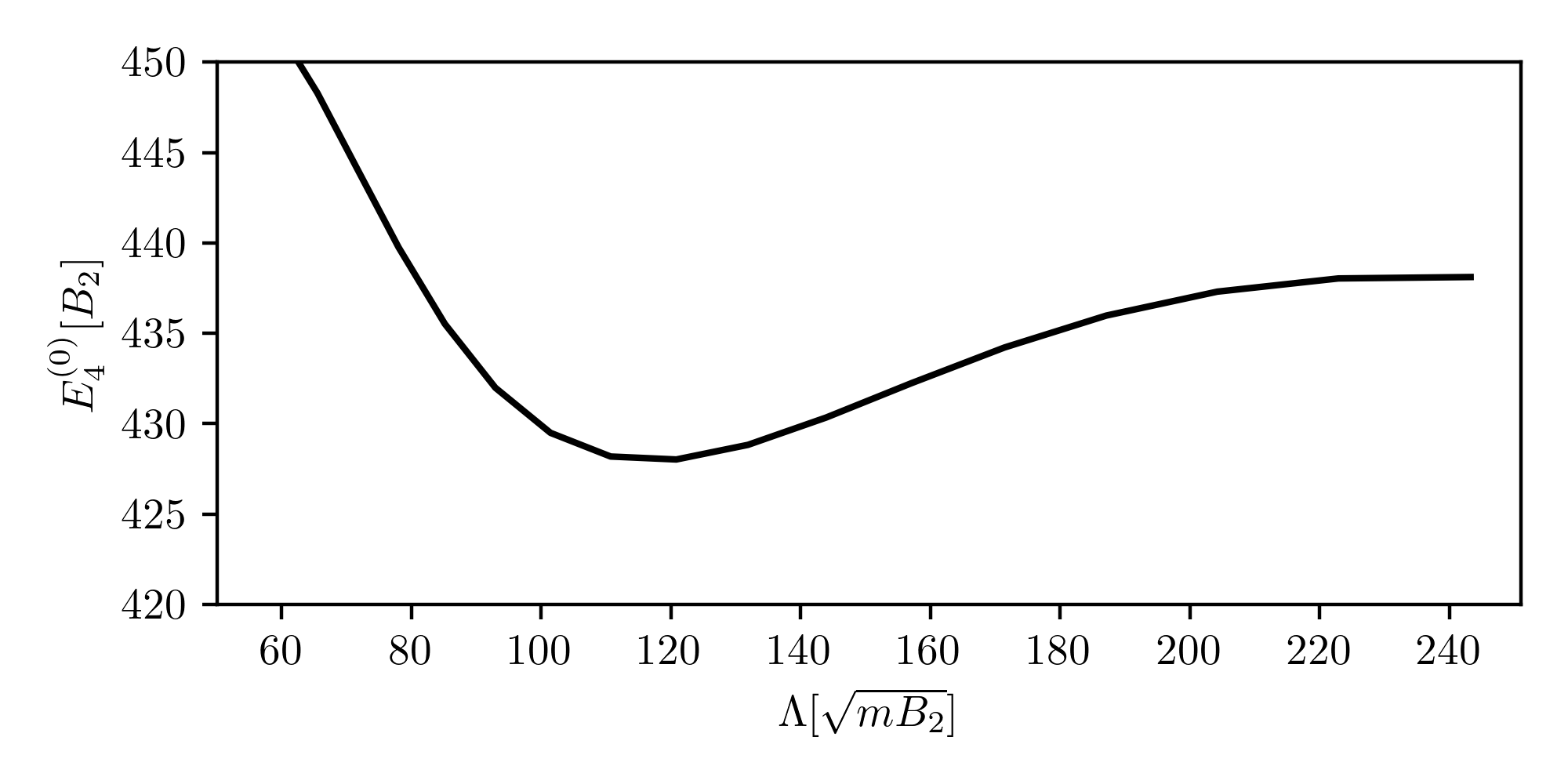}
\end{minipage}
\begin{minipage}[c]{\textwidth}
\centering    \includegraphics[width=10cm,trim={0.3cm 0 0 0},clip]{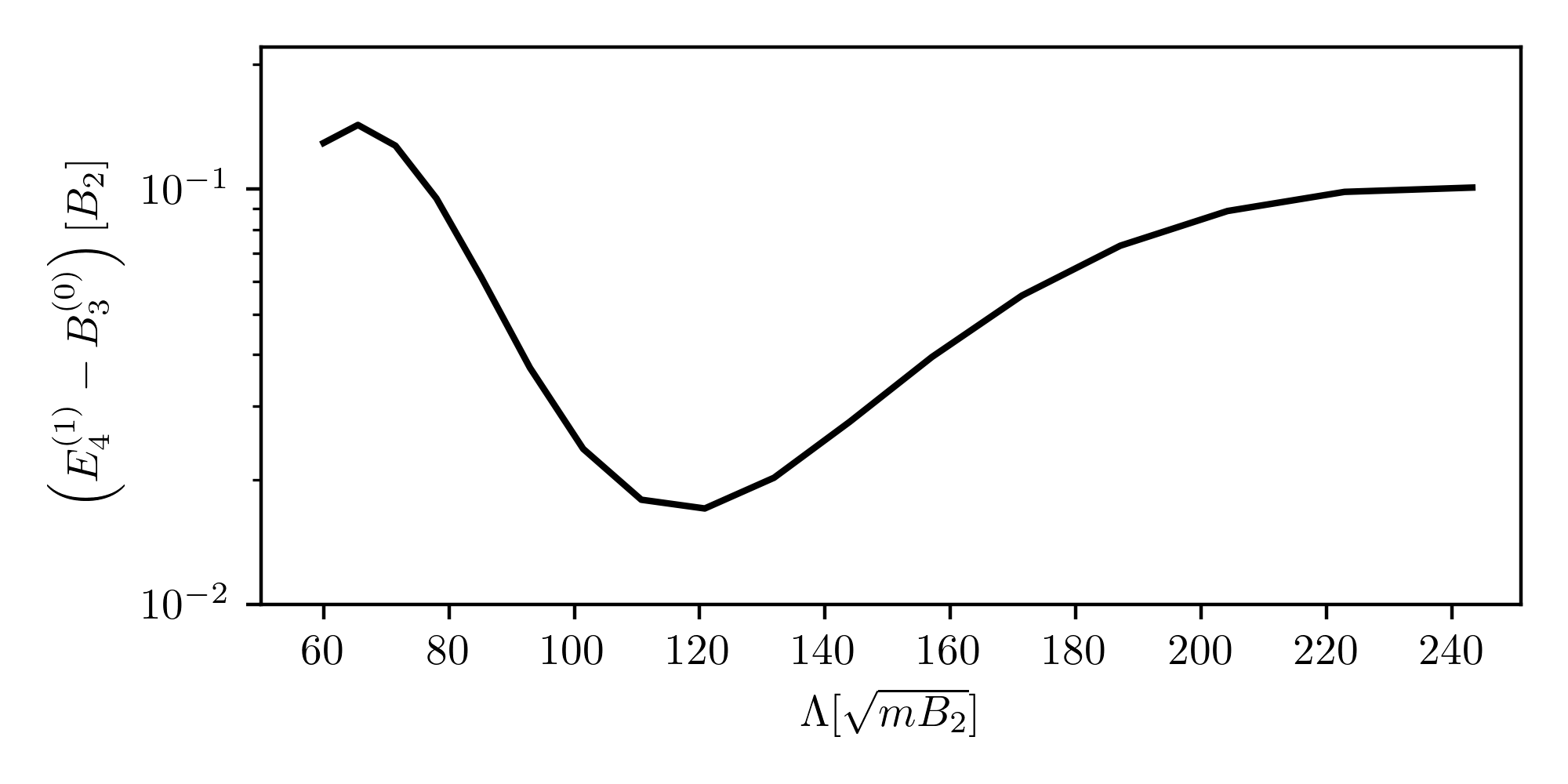}
\end{minipage}
\caption{Cutoff dependence of $E_4^{(0)}$(top) and  $E_4^{(1)}-B_3^{(0)}$ (bottom) at $\Lambda \lesssim 240\sqrt{mB_2}$. The data and notations are the same as in Fig.~\ref{fig: 4body-bindings}. For this range of $\Lambda$ there is no deep trimer and the complex tetramer binding energies $B_4^{(m)}$ are real and equal to $E_4^{(m)}$.}
\label{fig:Bslowcutoff}
\end{figure}

Fig.~\ref{fig: 4body-bindings} shows the cutoff dependence of $E_4^{(0)}$ on the top plot and $(E_4^{(1)} - B_3^{(0)})$ at the bottom for cold $^4$He atoms with the deep trimers either included or subtracted. The tetramer decay widths are small but non-zero when the deep trimers are included, while the widths are zero when the deep trimers are subtracted.
In both cases, $E_4^{(0)}$ and $(E_4^{(1)} - B_3^{(0)})$ converge at $\Lambda \gtrsim 1000\sqrt{mB_2}$. Note that the cutoff dependence in the range $250\sqrt{mB_2} \lesssim \Lambda \lesssim 1000\sqrt{mB_2}$ would be missed if the tetramer binding energies were only computed at $\Lambda\lesssim 240\sqrt{mB_2}$, as shown in Fig.~\ref{fig:Bslowcutoff}, and one could be misled and wrongly conclude that $E_4^{(m)}$ have already saturated at $\Lambda \approx 240\sqrt{mB_2}$.

\begin{table}[htb!]
    \centering
    \begin{tabular}{|c|c|c|c|c|c|}
    \hline
         &$B_3^{(1)}$~[mK] &$B_3^{(0)}$~[mK] &$E_4^{(1)}$~[mK]&$(E_4^{(1)}-B_3^{(0)})$~[mK] & $E_4^{(0)}$~[mK]  \\
    \hline
    Diagrammatic results &2.186&*128.5&128.517(1)&$0.017(1)$&526.1(5)\\
    \hline
    Platter et al.~\cite{Platter:2004he}    &*2.186&127&128[3]&1[3]&492[25]\\
    \hline
    Blume and Greene~\cite{Blume2000MonteCH}&2.186&125.5&132.7&7.2&559.7\\
    \hline
    Lazauskas and Carbonell~\cite{Lazauskas_2006}& 2.268&126.39&127.5&1.1&557.7\\
    \hline
    \end{tabular}
    \caption{Trimer and tetramer binding energies for cold $^4$He atoms with deep trimers subtracted, compared with Platter et al.~\cite{Platter:2004he}, Blume and Greene~\cite{Blume2000MonteCH}, and  Lazauskas and Carbonell~\cite{Lazauskas_2006}. The errors in the parentheses come from higher partial waves. The errors in the square brackets come from the residual cutoff dependence estimated by Platter et al.~\cite{Platter:2004he}. Both the diagrammatic results and the results of Platter et al.~\cite{Platter:2004he} have an additional EFT error of $\approx 10\%$, which is not shown explicitly here. The values with a star are used to fit the three-boson contact interaction. }
    \label{tab:Bs}
\end{table}
Table~\ref{tab:Bs} shows the diagrammatic results, compared with Platter et al.~\cite{Platter:2004he}, Blume and Greene~\cite{Blume2000MonteCH}, and Lazauskas and Carbonell~\cite{Lazauskas_2006}. The tetramer binding energies shown here are those with the deep trimers subtracted to keep the tetramers as bound states. Errors from higher partial waves are indicated in parentheses. Numerical errors are much smaller than those from higher partial waves and are not shown. 
While Platter et al.~\cite{Platter:2004he} did not include contributions from partial waves higher than S wave, they estimated their residual cutoff dependence to be  $\approx 2\%$ for $E_4^{(1)} $and $\approx 5\%$ for $E_4^{(0)}$, as shown in the square brackets in Table.~\ref{tab:Bs}.  Both the diagrammatic results and the results of Platter et al.~\cite{Platter:2004he} have an additional EFT error of $\approx 10\%$, which is not shown explicitly in Table~\ref{tab:Bs} because it is irrelevant for comparing two LO EFT calculations that should agree within numerical, momentum cutoff, and angular momentum cutoff errors.
The diagrammatic tetramer ground state ($E_4^{(0)}$) and excited state binding energy ($E_4^{(1)}$) agree with the other three calculations shown in Table~\ref{tab:Bs} within $\approx 6\%$ and $\approx 3\%$, respectively, which are consistent with the EFT error of $\approx 10\%$. However, a much smaller difference is found between $E_4^{(1)}$ and $B_3^{(0)}$, i.e., the
tetramer excited state in this calculation is extremely shallow, compared to what was found in the other three calculations.

\begin{figure}[htb!] 
  \label{ fig7} 
  \begin{minipage}[b]{0.5\linewidth}
    \centering
    \includegraphics[width=\linewidth]{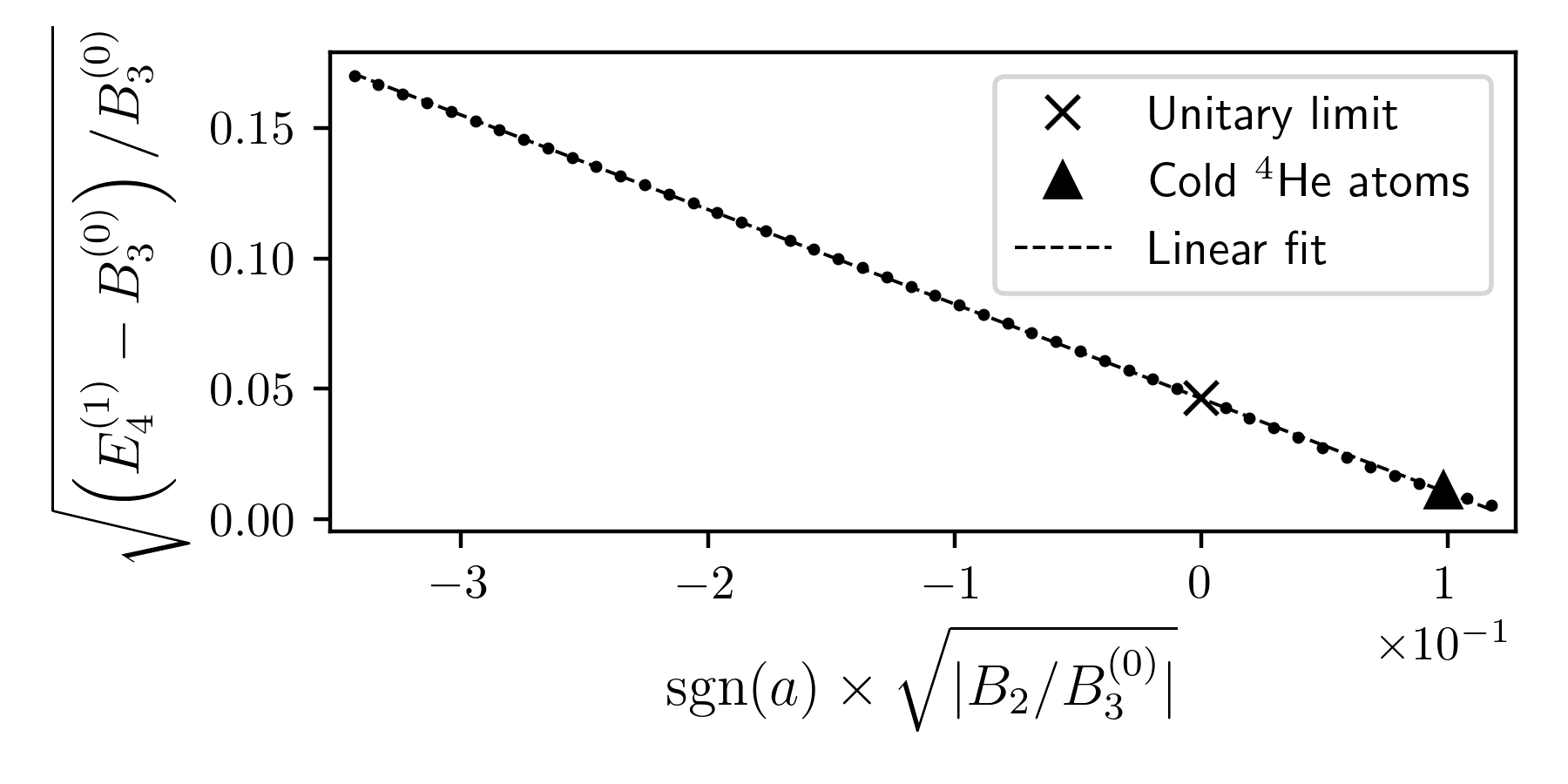} 
  \end{minipage}
  \begin{minipage}[b]{0.5\linewidth}
    \centering
    \includegraphics[width=\linewidth,trim={0 0.05cm 0  0 },clip]{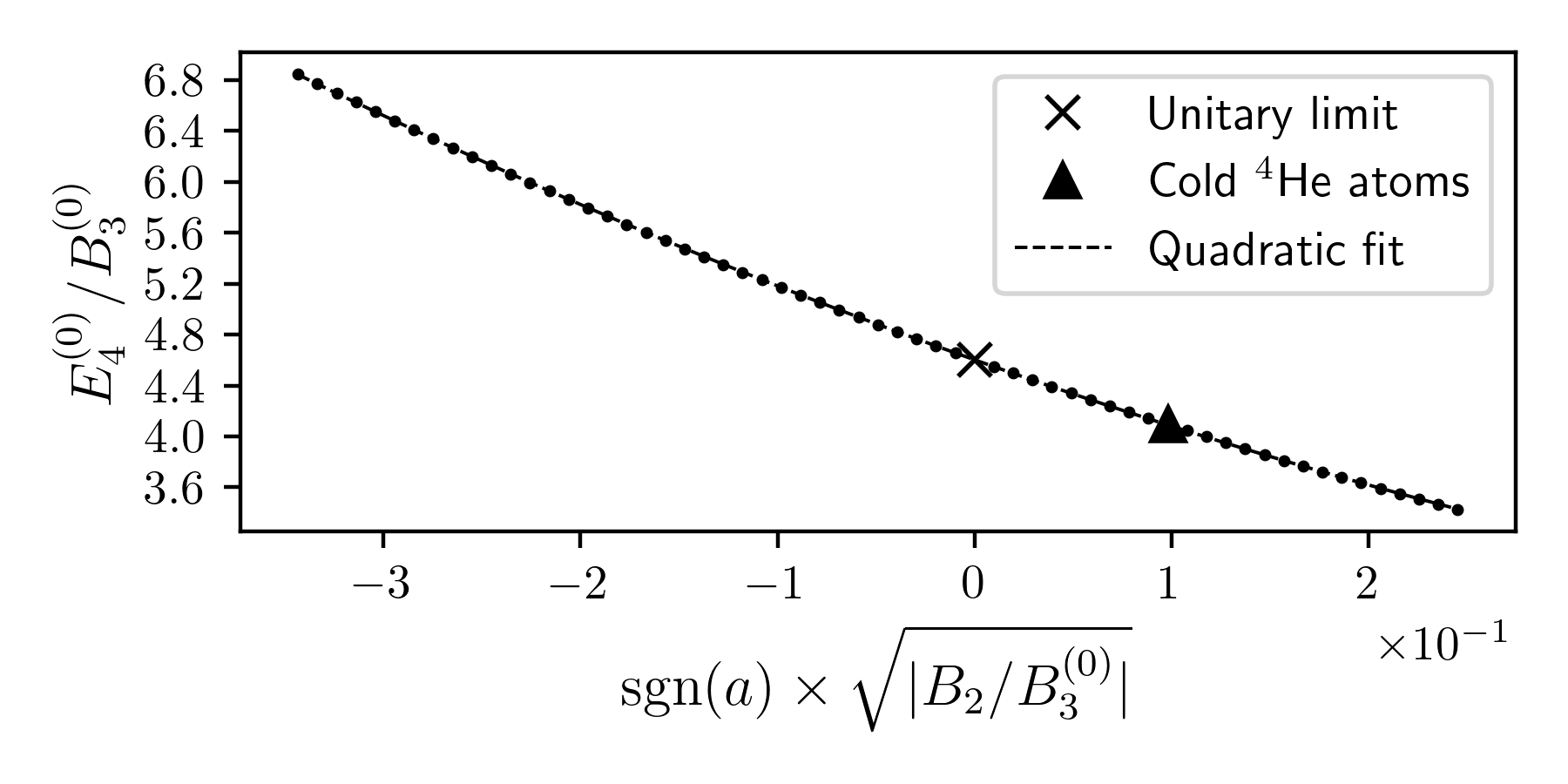} 
  \end{minipage} 
  \begin{minipage}[b]{0.5\linewidth}
    \centering
    \includegraphics[width=\linewidth]{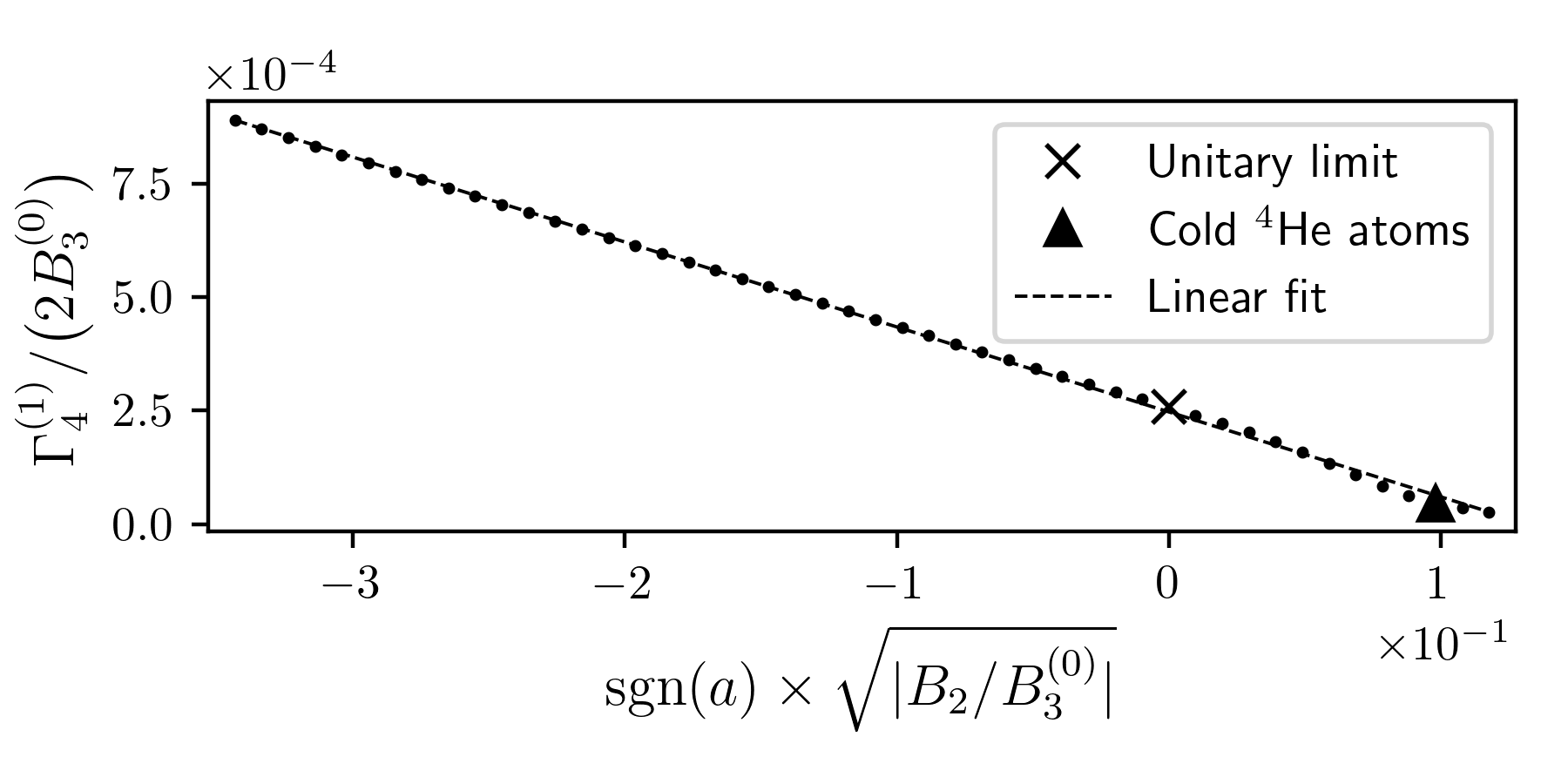} 
  \end{minipage}
  \begin{minipage}[b]{0.5\linewidth}
    \centering
    \includegraphics[width=\linewidth,trim={0 0.05cm 0  0 },clip]{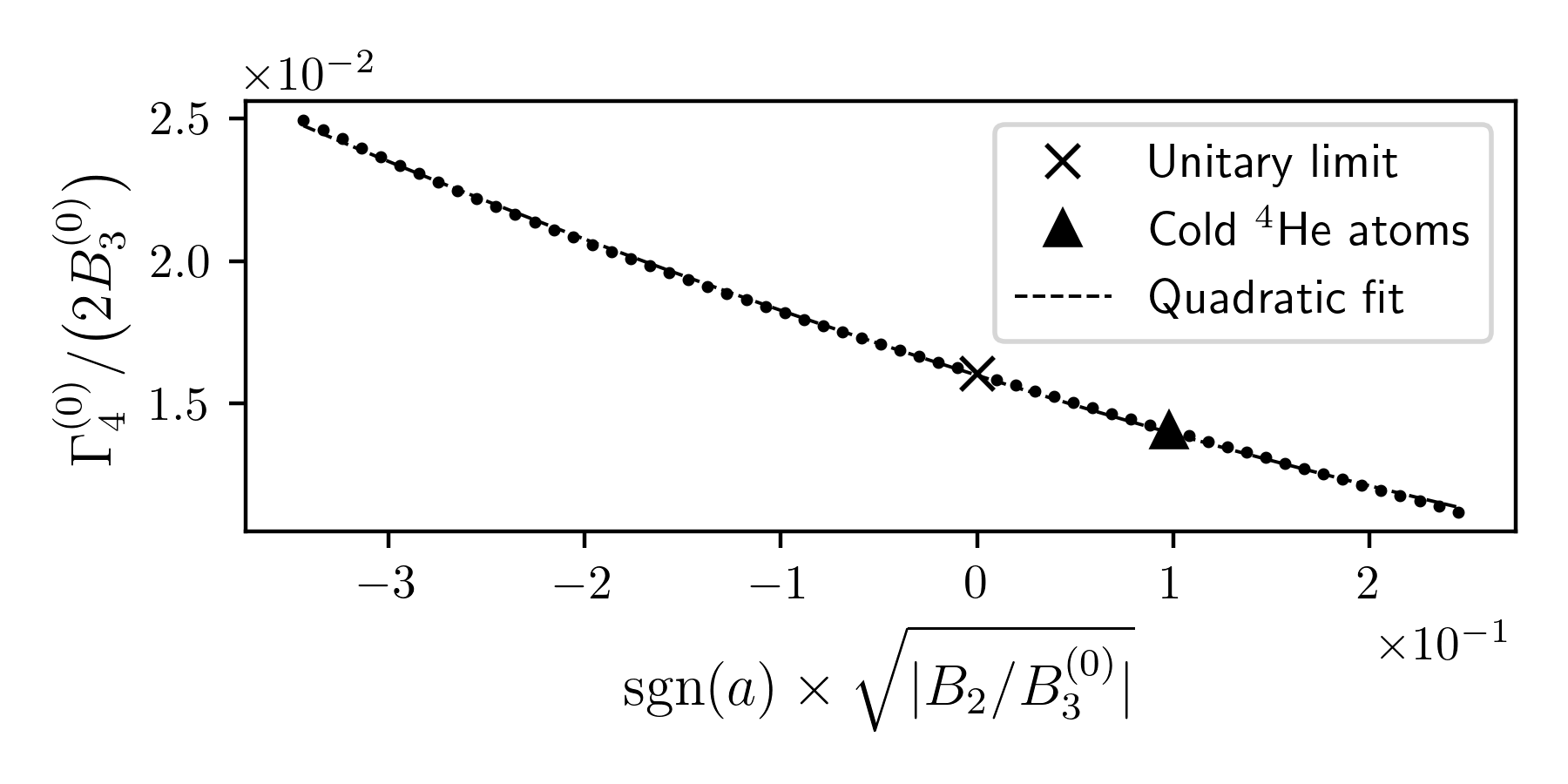} 
  \end{minipage}
\caption{Correlations between the tetramer and dimer binding energies (top), and between the tetramer decay width and the dimer binding energies (bottom) for the tetramer excited (left) and ground (right) states. The dots represent the numerical results from the diagrammatic approach evaluated at $\Lambda = 400\sqrt{mB_3^{(0)}}$, where one deep trimer state with $B_3^{(-1)}> E_4^{(0)}$ exists. The numerical results in the unitary (physical) limit are indicated by the cross (triangle). Dashed lines represent linear or quadratic fittings to the numerical results.}
\label{fig:B4Unitary}
\end{figure}

Fig.~\ref{fig:B4Unitary} shows the correlations between the tetramer binding energies (and decay widths) and dimer binding energies, evaluated at $\Lambda = 400\sqrt{mB_3^{(0)}}$ where one deep trimer state with $B_3^{(-1)}> E_4^{(0)}$ exists. 
For the tetramer excited state, the calculations are only performed for $B_2/B_3^{(0)}$ smaller than $1/103.9$, which is the value of $B_2/B_3^{(0)}$ in the diagrammatic calculation for cold $^4$He atoms. For a larger $B_2/B_3^{(0)}$, the binding energy of the tetramer excited state soon goes above the trimer-single-boson threshold for $B_3^{(0)}$. In the unitary limit ($B_2/B_3^{(0)}\to 0$), $\Gamma_4^{(0)}/B_3^{(0)}$ and $E_4^{(0)}/B_3^{(0)}$ are larger than their values in the physical limit for cold $^4$He atoms.\footnote{If in the numerical calculation there are not enough mesh points for discretized momenta close to $\sqrt{mB_2}$, the results for $\Gamma_4^{(0)}/B_3^{(0)}$ and $E_4^{(0)}/B_3^{(0)}$ will be misleadingly close to their values in the unitary limit due to numerical errors.}  Fittings (dashed lines) of the results in Fig.~\ref{fig:B4Unitary} yield
\begin{align}
    \sqrt{\frac{E_4^{(1)}-B_3^{(0)}}{B_3^{(0)}}}&=0.0463-0.361\times\textrm{sgn}(a)\sqrt{\abs{\frac{B_2}{B_3^{(0)}}}}, \nonumber\\
    \frac{E_4^{(0)}}{B_3^{(0)}} &=4.60 - 5.51\times\textrm{sgn}(a)\sqrt{\abs{\frac{B_2}{B_3^{(0)}}}} +  2.96\abs{\frac{B_2}{B_3^{(0)}}},
\end{align}
and 
\begin{align}
\frac{\Gamma_4^{(1)}}{2B_3^{(0)}} &=  
    2.47\times10^{-4} - 0.00187\times\textrm{sgn}(a)\sqrt{\abs{\frac{B_2}{B_3^{(0)}}}},\nonumber\\
    \frac{\Gamma_4^{(0)}}{2B_3^{(0)}} &=  
    0.0160 -0.0216\times\textrm{sgn}(a)\sqrt{\abs{\frac{B_2}{B_3^{(0)}}}}+  0.0112\abs{\frac{B_2}{B_3^{(0)}}}.
\end{align}

\begin{figure}[htb!]
    \centering
    \includegraphics[trim={0 0 0 0},clip,width=\textwidth]{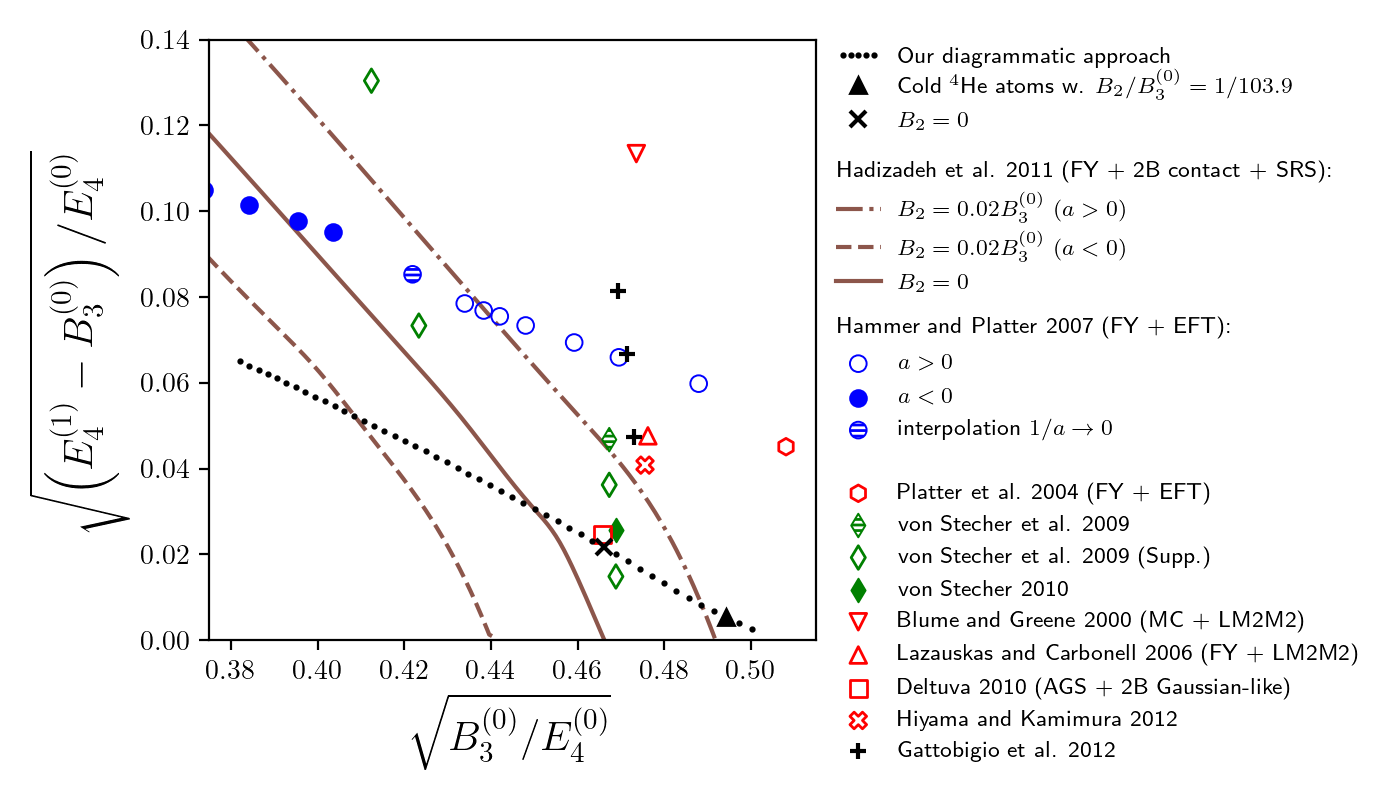}
    \caption{Correlations between the trimer and tetramer binding energies. Similar figures in Ref.~\cite{Hadizadeh2011,Frederico_2012} are supplemented with the diagrammatic results. The diagrammatic results plotted here are the same as those in Fig.~\ref{fig:B4Unitary} and are compared with previous calculations by Blume and Greene~\cite{Blume2000MonteCH}, Platter et al.~\cite{Platter:2004he}, Lazauskas et al.~\cite{Lazauskas_2006}, Hammer and Platter~\cite{Hammer_2007}, von Stecher et al.~\cite{2009NatPh...5..417V}, von Stecher~\cite{von_Stecher_2010}, Deltuva~\cite{Deltuva:2010xd},  Hadizadeh et al.~\cite{Hadizadeh2011}, Hiyama and Kamimura~\cite{Hiyama_2012}, and Gattobigio et al.~\cite{Gattobigio_2012}.}
    \label{fig:scalingFunction}
\end{figure}
The diagrammatic results around the unitary limit, along with other calculations from Refs.~\cite{2009NatPh...5..417V,von_Stecher_2010,Gattobigio_2012,Hammer_2007,Hiyama_2012,Blume2000MonteCH,Platter:2004he,Lazauskas_2006,Deltuva:2010xd,Hadizadeh2011}, are shown in Fig.~\ref{fig:scalingFunction}, where the correlation between $[(E_4^{(1)}-B_3^{(0)})/E_4^{(0)}]^{1/2}$  and $(B_3^{(0)}/E_4^{(0)})^{1/2}$ are plotted as was done in Refs.~\cite{Hadizadeh2011, Frederico_2012}. 
Hyperspherical and/or variational methods with various potential models are used by von Stecher et al.~\cite{2009NatPh...5..417V}, von Stecher~\cite{2009NatPh...5..417V}, Gattobigio et al.~\cite{Gattobigio_2012}, and Hiyama and Kamimura~\cite{Hiyama_2012}. Blume and Greene~\cite{Blume2000MonteCH} combined MC and the adiabatic hyperspherical approximation using the LM2M2 potential (MC+LM2M2); Platter et al.~\cite{Platter:2004he} solved the FY equation using EFT contact interactions (FY + EFT); Lazauskas and Carbonell~\cite{Lazauskas_2006} solved the FY equation using the LM2M2 potential (FY + LM2M2); Deltuva~\cite{Deltuva:2010xd} solved the AGS (Alt, Grassberger and Sandhas) equation~\cite{GRASSBERGER1967181}, which is equivalent to the FY equation, using a Gaussian-like two-body potential that also simulates many-body forces (AGS + 2B Gaussian-like); Hadizadeh et al.~\cite{Hadizadeh2011} solved the FY equation using a two-body contact interaction and the subtractive regularization scheme~\cite{Adhikari1995} for three- and four-boson propagators (FY + 2B contact + SRS). Most of the data shown in Fig.~\ref{fig:scalingFunction} are compiled in Ref.~\cite{Hadizadeh_2012,Frederico_2012,HadizadehPC}. Each curve for Hadizadeh et al.~\cite{Hadizadeh2011} shown in Fig.~\ref{fig:scalingFunction} uses a fixed value of $B_2/B_3^{(0)}$ and is driven by a four-body scale introduced in their calculations, while their different curves use different values of $B_2/B_3^{(0)}$. This is in contrast to Hammer and Platter~\cite{Hammer_2007} and the diagrammatic calculation in this paper, where no four-body scale is introduced and each value of $B_2/B_3^{(0)}$ corresponds to a single point (as opposed to a curve) in Fig.~\ref{fig:scalingFunction}.
The diagrammatic result in the unitary limit, represented by ``$\times$", as well as the three points nearest to it from other calculations in Fig.~\ref{fig:scalingFunction} are also shown in Table~\ref{tab:unitaryBs}, where the uncertainties for the diagrammatic results, shown in the parenthesis, are from higher partial waves.
\begin{table}[htb!]
    \centering
    \begin{tabular}{|c|c|c|c|c|}
    \hline
         &$E_4^{(0)}/B_3^{(0)}$ & $\Gamma_4^{(0)}/(2B_3^{(0)})$ & $E_4^{(1)}/B_3^{(0)}$& $\Gamma_4^{(1)}/(2B_3^{(0)})$\\
    \hline
    Diagrammatic results       &4.60(1)&0.0160(1)& 1.0022(3)& $2.57(2)\times 10^{-4}$\\
    \hline
    Deltuva \cite{Deltuva:2010xd}   &4.6108&0.01484&1.00228&$2.38\times 10^{-4}$\\
    \hline
    von Stecher~\cite{von_Stecher_2010}  &4.55&-&1.003&-\\
    \hline
    von Stecher et al. (Supp., with $V_{a}$ and $V_{3b}$)~\cite{2009NatPh...5..417V} &4.55&-&1.001&-\\
    \hline
    \end{tabular}
    \caption{Tetramer binding energies and decay widths in or close to the unitary limit. The first row shows the diagrammatic results obtained with $B_2/B_3^{(0)}=0$. The numbers in the other three rows correspond to the three data points closest to the diagrammatic result in the unitary limit among all the points in Fig.~\ref{fig:scalingFunction}. The numbers in the last row were presented in the supplementary information of von Stecher et al.~\cite{2009NatPh...5..417V} using a Gaussian two-body potential ($V_a$) and a Gaussian three-body potential ($V_{3b}$).}
    \label{tab:unitaryBs}
\end{table}
The diagrammatic tetramer binding energies $E_4^{(1)}/B_3^{(0)}$  and $E_4^{(0)}/B_3^{(0)}$ align with the other three calculations shown in Table~\ref{tab:unitaryBs}. Moreover, the diagrammatic results of $\Gamma_4^{(0)}/(2B_3^{(0)})$ and $\Gamma_4^{(1)}/(2B_3^{(0)})$ show great agreement with Deltuva~\cite{Deltuva:2010xd} while only Deltuva~\cite{Deltuva:2010xd} and this paper have computed the tetramer decay widths in the unitary limit among the calculations shown in Fig.~\ref{fig:scalingFunction}.

Another noteworthy point in Fig.~\ref{fig:scalingFunction} is that, as suggested by the data from the diagrammatic calculation and Hammer and Platter~\cite{Hammer_2007}, $(E_4^{(1)} - B_3^{(0)})/E_4^{(0)}$ decreases and $B_3^{(0)}/E_4^{(0)}$ increases when $B_2/B_3^{(0)}$ flows from the unitary limit to the physical limit of cold $^4$He atoms with $a>0$. In fact, the diagrammatic result of $E_4^{(1)} /B_3^{(0)} = 1.0022(3)$ in the unitary limit is very close to that for cold $^4$He atoms, $E_4^{(1)}/B_3^{(0)} = 1.00013(1)$. This is not surprising because, as already known, the system of cold $^4$He atoms with a large atom-atom scattering length $a_{^4\textrm{He}}\gg l_{^4\textrm{He}}$ is close to the unitary limit. Furthermore, while $E_4^{(1)}/B_3^{(0)}$ decreases from the unitary limit to the physical limit of cold $^4$He atoms with $a>0$, the beginning diagrammatic value of $E_4^{(1)}/B_3^{(0)} = 1.0022(3)$ in the unitary limit is much closer to one, compared to, for example, $E_4^{(1)}/B_3^{(0)} = 1.01$ found by Hammer and Platter~\cite{Hammer_2007}. This feature carries over to the physical limit of cold $^4$He atoms and partially explains why for cold $^4$He atoms the diagrammatic result of $E_4^{(1)}$ is much closer to $E_3^{(0)}$ than the other calculations in Table~\ref{tab:Bs}. An NLO EFT calculation in the future can also help to better understand the tetramer excited state. 

\section{Summary and Outlook}
\label{sec:conclusions}
This paper examined the cutoff dependence of tetramer binding energies for cold $^4$He atoms at large cutoffs where one or more deep trimers exist. The four-boson calculation in this paper is a gateway to NLO and higher order calculations and nuclear systems with four or more bodies. A four-boson integral equation was constructed via Feynman diagrams using an EFT with two- and three-boson contact interactions. This integral equation was written in terms of three-boson amplitudes, whose poles and residues were calculated and then included or subtracted through the Cauchy principal value prescription. In this way, the tetramer binding energies at cutoffs above $\Lambda_t$ were calculated for the first time. In particular, while reaffirming that no four-boson contact interaction is required at LO, the diagrammatic results also demonstrated the necessity of going above $\Lambda_t$ for the tetramer binding energies to converge. Furthermore, the converged diagrammatic results do not have errors from the residual cutoff dependence, which can be sizeable compared to the EFT error at NLO or higher orders. 

For cold $^4$He atoms, a tetramer ground state binding energy of 526.1(5)~mK and a tetramer excited state only $0.017(1)$~mK below the trimer ground state with a binding energy $B_3^{(1)} = 128.5$~mK are obtained. These diagrammatic results have converged sufficiently as functions of cutoff and agree with previous calculations~\cite{Platter:2004he,Lazauskas_2006,Blume2000MonteCH}. 
In the unitary limit, the two tetramers associated with the trimer with a (arbitrary) binding energy $B_3^{(0)}$ by setting $B_2 = 0$ fitting the three-boson force to $B_3^{(0)}$ at all cutoffs. For the tetramer ground state, a binding energy $E_4^{(0)}/B_3^{(0)} = 4.60(1)$ and decay width $\Gamma_4^{(0)}/(2B_3^{(0)}) = 0.0160(1)$ are obtained. For the tetramer excited state, $E_4^{(1)}/B_3^{(0)} = 1.0022(3)$ and  $\Gamma_4^{(1)}/(2B_3^{(0)}) = 2.57(2)\times10^{-4}$ are obtained. These tetramer binding energies and decay widths in the unitary limit show good agreement with previous studies~\cite{von_Stecher_2010,2009NatPh...5..417V,Deltuva:2010xd}. 

The four-body diagrammatic approach with trimer poles addressed provides several possible directions for future four-body calculations. First, while Ref.~\cite{Bazak_2019} found the need for a four-boson force at NLO to maintain the RG invariance of the tetramer binding energies, their calculation is limited at relatively low cutoffs where no deeply bound trimer state exists. Using the integral equation developed in this paper, it is possible to perform an NLO four-boson calculation at higher cutoffs.   Moreover, instead of fitting the four-boson force to $E_4^{(0)}$ as was done in Ref.~\cite{Bazak_2019}, it would also be interesting to fit it to $E_4^{(1)}$. One motivation for this is that, even though the values of $E_4^{(1)}$ found in different calculations are close to each other, the differences between $E_4^{(1)}$ and $B_3^{(0)}$ vary significantly, as shown in Table~\ref{tab:Bs} and Fig.~\ref{fig:scalingFunction}. This may be due to different values for the two-body range correction, which is zero in a LO EFT calculation. (See, e.g., Refs.~\cite{Frederico_2012, D_Incao_2018, frederico2023universal} for discussions on the effect of range corrections.)  An NLO EFT calculation at large cutoffs with minimized errors from cutoff variations can provide further insights into the effect of a non-zero effective range and four-boson force on the tetramer excited state, trimer-single-boson scattering length, and possibly the tetramer limit cycle. Second, while this paper focuses on tetramer binding energies, the technique of addressing trimer poles can also be applied to study scatterings. One can then obtain the four-body contribution to the loss rate of trapped atoms.
Last, the diagrammatic four-boson integral equation can be extended to four-nucleon systems by including spin and isospin degrees of freedom. Four-nucleon calculations can be performed at large cutoffs by subtracting deep three-nucleon poles using the method described in this paper. For example, one may use \eftnopi to perform form-factor-like calculations in four nucleon systems at cutoffs above $2000$~MeV and benchmark previous potential-model calculations and effective-theory calculations at lower cutoffs.

\appendix
\acknowledgments{I would like to thank Shailesh Chandrasekharan, Sebastian König, Son Nguyen, Lucas Platter, Roxanne Springer, Jared Vanasse, and Feng Wu for useful discussions and feedback. I would also like to thank Mohammadreza Hadizadeh for sharing their collaboration's data. This work is supported by the Henry W. Newson fellowship for Fall 2022 and by the U.S. Department of Energy, Office of Science, Office of Nuclear Physics, under Award Number DE-FG02-05ER41368.}
\section{Other examples of four-boson diagrams}
\label{app:baddiagrams}
\begin{figure}[htb!]
    \centering    \includegraphics[trim={20cm 7cm 20cm 10cm},clip,width=7in]{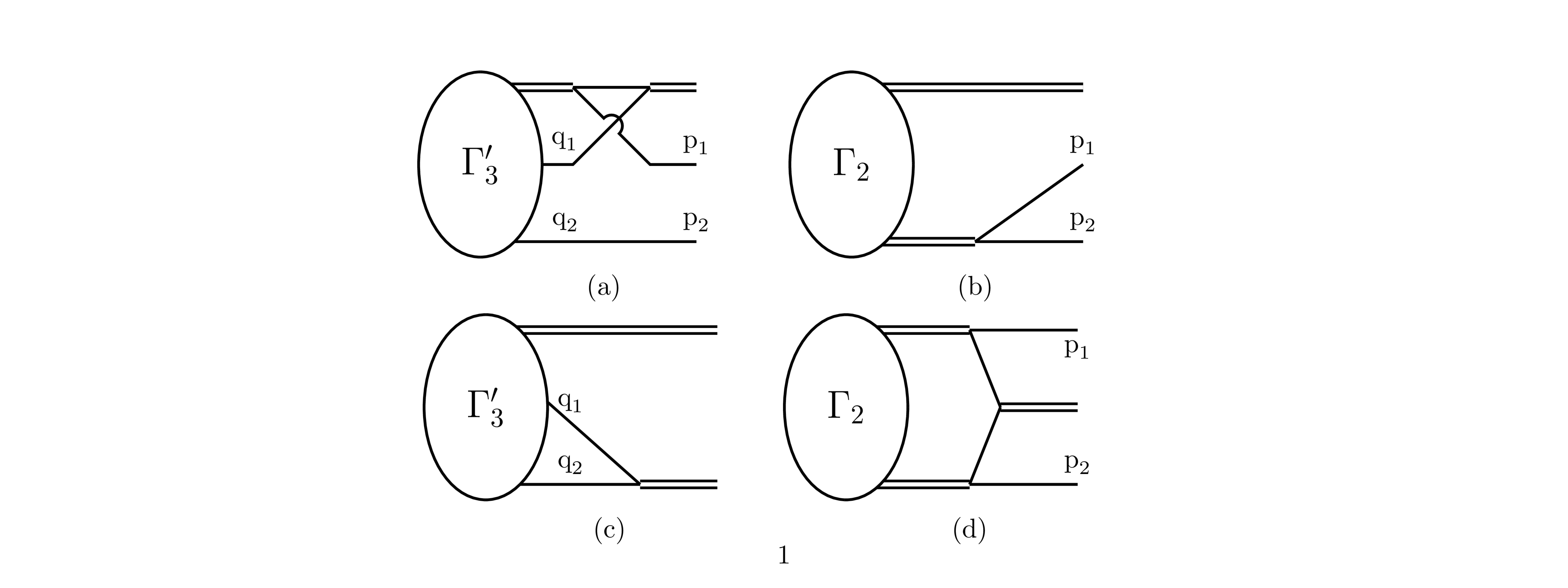}
    \caption{Examples of four-boson Feynman diagrams. }
    \label{fig:examples of 4body diagrams}
\end{figure}
Four examples of four-boson Feynman diagrams are shown in Fig.~\ref{fig:examples of 4body diagrams}. 
For these diagrams, $\textrm{p}_i = \{E_{p_i}, \vect{p}_i\}$ and $\textrm{q}_i = \{E_{q_i}, \vect{q}_i\}$ are the off-shell four-momenta of the single-boson states in the four-boson CM frame.
Although it may be tempting to construct a four-boson integral equation using diagrams (a), (b), and (c) in Fig.~\ref{fig:examples of 4body diagrams} as the kernel, there are two difficulties. First, iterations over diagrams (b) and (c) in Fig.~\ref{fig:examples of 4body diagrams} will overcount the number of the two-boson loops. Second, an integral equation that is closed under on-shell $\textrm{p}_1$ and $\textrm{p}_2$ is preferred to minimize the number of integration variables. This, however, cannot be constructed directly using these three diagrams as the kernel of the integral equation.  To see this, consider diagram (d) which is obtained by attaching diagram (a) without $\Gamma_3$ to the right side of diagram (b). (Diagram (d) is part of the kernel of the integral equation shown in Fig.~\ref{fig:Intg_K3}.) The energy loop integral in diagram (d) has a branch cut from each dimer propagator attached to $\Gamma_2$, one in the upper half and the other in the lower half of the complex energy plane, which makes it difficult to invoke the residue theorem on this loop energy integral. Thus, one of the two nucleon lines in diagram (b) is generally off-shell.

\section{Four-boson integral equation in operator form}
\label{app:four-body}
\setcounter{equation}{0}
\renewcommand{\theequation}{\thesection\arabic{equation}}
This appendix shows the derivation of Eqs.~\eqref{eq:formal_Intg_T3} and~\eqref{eq:gamma2_main} from Eq.~\eqref{eq:formal_Intg_K3}. Eq.~\eqref{eq:formal_Intg_K3} reads
\begin{equation}
    \begin{aligned}
    \Gamma'_3& =(1 + \mathcal{P}_3) \left( K_{33}\Gamma'_3 + K_{32} \Gamma_2\right)\\
     \Gamma_2& = (1 + \mathcal{P}_2) \left(K_{23} \Gamma'_3 + K_{22} \Gamma_2\right).
\end{aligned}
\label{eq:intg_K3_app}
\end{equation}
The first line can be used to write $\Gamma'_3$ in terms of $\Gamma_2$:
\begin{align}
    \Gamma'_3& = \left(\mathrm{1} - (1 + \mathcal{P}_3)K_{33} \right)^{-1}(1 + \mathcal{P}_3)K_{32}\Gamma_2,
    \label{eq:gamma3sub}
\end{align}
which can be plugged into the second line in Eq.~\eqref{eq:intg_K3_app}, yielding
\begin{align}
     \Gamma_2 = (1 + \mathcal{P}_2)\left[ K_{23} \left(\mathrm{1} - (1 + \mathcal{P}_3)K_{33} \right)^{-1}(1 + \mathcal{P}_3)K_{32} + K_{22} \right]\Gamma_2
     \label{eq:gamma2}
\end{align}
as long as $(\mathrm{1} - (1 + \mathcal{P}_3)K_{33})$ is invertible at the energy under consideration. To further rewrite the equation two identities are needed. The first one is 
\begin{align}
     \left(\mathrm{1} - (1 + \mathcal{P}_3)K_{33} \right)^{-1} &=  \left(\left(1 -  \mathcal{P}_3K_{33}\left(\mathrm{1} - K_{33}\right)^{-1}\right)\left(\mathrm{1} - K_{33}\right)\right) ^{-1} \nonumber\\
     &=  \left(\mathrm{1} - K_{33}\right)^{-1}
     \left(1 -  \mathcal{P}_3K_{33}\left(\mathrm{1} - K_{33}\right)^{-1}\right)^{-1} \nonumber \\
      &=  \left(\mathrm{1} + T_{33}\right)
     \left(1 -  \mathcal{P}_3T_{33}\right)^{-1}
     \label{eq:K33_modified_1}
\end{align}
where
\begin{align}
    T_{33} \equiv K_{33}(1-K_{33})^{-1} 
\end{align}
The second one is
\begin{align}
    \left(1 -  \mathcal{P}_3T_{33}\right)^{-1}(1 + \mathcal{P}_3) &= \left(1 + \left(1 -  \mathcal{P}_3T_{33}\right)^{-1}\mathcal{P}_3T_{33}\right) +  \left(1 -  \mathcal{P}_3T_{33}\right)^{-1}\mathcal{P}_3\nonumber\\
    &=1 + \left(1 -  \mathcal{P}_3T_{33}\right)^{-1}\mathcal{P}_3(1 + T_{33})
    \label{eq:K33_modified_2}
\end{align}
Using Eqs.~\eqref{eq:K33_modified_1} and ~\eqref{eq:K33_modified_2}, one can write Eq.~\eqref{eq:gamma2} as
\begin{align}
    \Gamma_2 &= (1 + \mathcal{P}_2)\left[ K_{23} \left(\mathrm{1} + T_{33}\right)
     \left(1 -  \mathcal{P}_3T_{33}\right)^{-1}(1 + \mathcal{P}_3)K_{32} + K_{22} \right]\Gamma_2\nonumber\\
      &= (1 + \mathcal{P}_2)
      \Bigl[ K_{23} \left(\mathrm{1} + T_{33}\right)
     \left(1 -  \mathcal{P}_3T_{33}\right)^{-1}\mathcal{P}_3(1 + T_{33}) K_{32} \nonumber \\
     & \hspace{2cm}+  K_{23} \left(\mathrm{1} + T_{33}\right)K_{32} + K_{22} \Bigr]\Gamma_2\nonumber\\
     & = (1 + \mathcal{P}_2)\left[ K'_{23}\left(1 -  K'_{33}\right)^{-1}K'_{32} + K'_{22}\right]\Gamma_2 
     \label{eq:gamma2_new}
\end{align}
where
\begin{equation}
    \begin{aligned}
    K'_{23} &\equiv K_{23}\left(\mathrm{1} + T_{33}\right)  \\
    K'_{32} &\equiv \mathcal{P}_3\left(\mathrm{1} + T_{33}\right)K_{32}\\
    K'_{22} &\equiv K_{23}\left(\mathrm{1} + T_{33}\right)K_{32}+ K_{22}\\
    K'_{33} &\equiv \mathcal{P}_3T_{33}
\end{aligned}
\label{eq:K's_app}
\end{equation}
as claimed in Eq.~\eqref{eq:K's}. Note that Eq.~\eqref{eq:gamma2_new} is just Eq.~\eqref{eq:gamma2_main} and has the same form as Eq.~\eqref{eq:gamma2}. Similarly, one can use Eqs.~\eqref{eq:K33_modified_1} and~\eqref{eq:K33_modified_2} to rewrite Eq.~\eqref{eq:gamma3sub} in terms of the quantities defined in Eq.~\eqref{eq:K's_app}:
\begin{align}
    \Gamma'_3 &= (1+\mathcal{P}_3)\left(1-K'_{33}\right)^{-1}K'_{32}\Gamma_2\\
    &\equiv(1+\mathcal{P}_3)\Gamma_3
\end{align}
where $\Gamma_3$ satisfy
\begin{align}
    \Gamma_3 &= K'_{33}\Gamma_3 + K'_{32}\Gamma_2.
    \label{eq:gamma3prime}
\end{align}
Combining Eqs.~\eqref{eq:gamma3prime} and~\eqref{eq:gamma2_new} gives
\begin{align}
    \Gamma_3 &= K'_{33}\Gamma_3 + K'_{32}\Gamma_2\nonumber\\
    \Gamma_2 &= (1+\mathcal{P}_2)(K'_{23}\Gamma_3 + K'_{22}\Gamma_2),
\end{align}
as claimed in Eq.~\eqref{eq:formal_Intg_T3}. 

\section{Momentum and partial-wave basis, and change of basis}
\label{app: identities}
For a four-boson system with zero total angular momentum, states in the angular momentum basis are given by
\begin{align}
    |L = 0, m_L = 0, (\ell\lambda)k_1k_2\rangle&=\int\!\!\frac{d\Omega_{k_1}}{4\pi}\int\!\!\frac{d\Omega_{k_2}}{4\pi}\sum_{m,\rho}\delta_{\ell, \lambda}\delta_{m, -\rho}\frac{(-1)^{\lambda - \rho}}{\sqrt{2\ell + 1}}Y_\ell^m(\widehat{\boldsymbol{k}}_1)Y_\lambda^\rho(\widehat{\boldsymbol{k}}_2)|\vect{k}_1\vect{k}_2\rangle\nonumber\\
    &=\int\!\!\frac{d\Omega_{k_1}}{4\pi}\int\!\!\frac{d\Omega_{k_2}}{4\pi}\sum_{m,\rho}\delta_{\ell, \lambda}\delta_{m, -\rho}\frac{(-1)^{\lambda }}{\sqrt{2\ell + 1}}Y_\ell^m(\widehat{\boldsymbol{k}}_1)Y_\ell^{-\rho*}(\widehat{\boldsymbol{k}}_2)|\vect{k}_1\vect{k}_2\rangle\nonumber\\
     &=\int\!\!\frac{d\Omega_{k_1}}{4\pi}\int\!\!\frac{d\Omega_{k_2}}{4\pi}\delta_{\ell, \lambda}(-1)^{\lambda }\sqrt{2\ell + 1}P_l\left(\widehat{\boldsymbol{k}}_1\cdot \widehat{\boldsymbol{k}}_2\right)|\vect{k}_1\vect{k}_2\rangle,
\end{align}
where the addition theorem of spherical harmonics is used on the last line. The overlap between four-body states under different momentum bases (i.e., Jacobi or \fourBodyMom) is given by
\begin{align}
    &\langle (\ell\lambda) p_1 p_2 | (\ell'\lambda') p^J_1 p^J_2\rangle\nonumber\\
    =&
     \delta_{\ell, \lambda}\delta_{\ell', \lambda'}(-1)^{\ell + \ell' }\sqrt{(2\ell + 1)(2\ell' + 1)} \frac{2\pi^2}{(q_2)^2}\delta\left(q_2 - p_2\right) \nonumber\\[-0.2em]
    &
   \times \int \frac{d\Omega_{p_1}}{4\pi}\frac{d\Omega_{p_2}}{4\pi}
   P_\ell\left(\hatvect{p}_1\cdot\hatvect{p}_2\right)
   P_{\ell'}\left(\hatvect{p}^J_1(\vect{p}_1,\vect{p}_2)\cdot\hatvect{p}_2^J(\vect{p}_2)\right).
\end{align}

The matrix element of $T_{33}$ under \fourBodyMom is related to $ T_{33,~\textrm{JJ}}^{(\ell\lambda; \ell'\lambda')}\left(p_1^J, p_2^J; q_1^J,q_2^J\right)$, given by Eq.~\eqref{eq:T33_Jacobi}, through a change of basis: 
\begin{align}
    &\quad T_{33,~\textrm{SJ}}^{(\ell\lambda; \ell'\lambda')}\left(p_1, p_2; q_1^J,q_2^J\right)\nonumber\\
    &\equiv
    \langle (\ell\lambda) p_1 p_2 |T_{33}| (\ell'\lambda') q_1^J q_2^J\rangle \nonumber \\[0.5em]
    &\begin{aligned}
    =\sum_{\tilde{\ell},\tilde{\lambda}}&\iint_0^\Lambda  \frac{(p^J_1)^2dp^J_1}{2\pi^2}\frac{(p^J_2)^2dp^J_2}{2\pi^2} \langle (\ell\lambda) p_1 p_2 | (\tilde{\ell}\tilde{\lambda}) p^J_1 p^J_2\rangle T_{33,~\textrm{JJ}}^{(\widetilde{\ell}\widetilde{\lambda}; \ell'\lambda')}\left(p_1^J, p_2^J; q_1^J,q_2^J\right)
    \end{aligned}\nonumber \\[0.5em]
    &\begin{aligned}
    =&~
     \delta_{\ell, \lambda}\delta_{\ell', \lambda'}(-1)^{\ell + \ell' }\sqrt{(2\ell + 1)(2\ell' + 1)} \frac{2\pi^2}{(p_2)^2}\delta\left(p_2 - q_2^J\right) \\[-0.2em]
    &
   \times \iint \frac{d\Omega_{p_1}}{4\pi}\frac{d\Omega_{p_2}}{4\pi}
   P_\ell\left(\hatvect{p}_1\cdot\hatvect{p}_2\right)
   P_{\ell'}\left(\hatvect{p}^J_1(\vect{p}_1,\vect{p}_2)\cdot\hatvect{p}_2^J(\vect{p}_2)\right) \widetilde{T}_{33,~\textrm{JJ}}^{\ell'}\left(p_1^J(\vect{p}_1, \vect{p}_2); q_1^J,q_2^J\right)
    \end{aligned}\nonumber \\[0.5em]
    &\begin{aligned}
    \equiv&~
     \delta_{\ell, \lambda}\delta_{\ell', \lambda'}(-1)^{\ell + \ell' }\sqrt{(2\ell + 1)(2\ell' + 1)} \frac{2\pi^2}{(p_2)^2}\delta\left(p_2 - q_2^J\right) \widetilde{T}_{33,~\textrm{SJ}}^{(\ell;\ell')}\left(p_1,p_2; q_1^J\right),
    \end{aligned}
    \label{eq:T_33_NJ}
\end{align}
where $\widetilde{T}_{33,~\textrm{SJ}}^{\ell;\ell'}\left(p_1; q_1^J,q_2^J\right)$ is defined implicitly on the last line. $\widetilde{T}_{33,~\textrm{JS}}^{\ell'}\left(p_1^J; q_1,q_2\right)$ can be defined in a similar manner:
\begin{align}
    &\widetilde{T}_{33,~\textrm{JS}}^{\ell;\ell'}\left(p_1^J; q_1,q_2\right)\nonumber\\
    =&\iint \frac{d\Omega_{q_1}}{4\pi}\frac{d\Omega_{q_2}}{4\pi}
    P_{\ell}\left(\hatvect{p}^J_1\cdot\hatvect{q}_2\right) 
   P_{\ell'}\left(\hatvect{q}_1^J(\vect{q}_1,\vect{q}_2)\cdot\hatvect{q}_2^J(\vect{q}_2)\right)
   \widetilde{T}_{33,~\textrm{JJ}}^{\ell'}\left(p_1^J; q_1^J(\vect{q}_1,\vect{q}_2),q_2^J(\vect{q}_2)\right).
   \label{eq:Ttilde_33_JN}
\end{align}
The matrix element of $T_{33}$ in \NormPW is related to $T_{33,~\textrm{JJ}}^{(\ell\lambda; \ell'\lambda')}\left(p_1^J, p_2^J; q_1^J,q_2^J\right)$ by a change of basis:
\begin{align}
    &\quad T_{33,~\textrm{SS}}^{(\ell\lambda; \ell'\lambda')}\left(p_1, p_2; q_1,q_2\right)\nonumber\\
    &\equiv
    \langle (\ell\lambda) p_1 p_2 |T_{33}| (\ell'\lambda') q_1 q_2\rangle \nonumber \\[0.5em]
    &\begin{aligned}
    =\sum_{\tilde{\ell},\tilde{\lambda},\tilde{\ell}',\tilde{\lambda}'}&\iiiint_0^\Lambda  \frac{(p^J_1)^2dp^J_1}{2\pi^2}\frac{(p^J_2)^2dp^J_2}{2\pi^2}\frac{(q^J_1)^2dq^J_1}{2\pi^2}\frac{(q^J_2)^2dq^J_2}{2\pi^2}\\[-0.2em]
    &
    \times \langle (\ell\lambda) p_1 p_2 | (\tilde{\ell}\tilde{\lambda}) p^J_1 p^J_2\rangle \langle  (\tilde{\ell}'\tilde{\lambda}') q^J_1 q^J_2 | (\ell'\lambda') q_1 q_2\rangle T_{33,~\textrm{JJ}}^{(\widetilde{\ell}\widetilde{\lambda}; \widetilde{\ell}'\widetilde{\lambda}')}\left(p_1^J, p_2^J; q_1^J,q_2^J\right)
    \end{aligned}\nonumber \\[0.5em]
    &\begin{aligned}
    =&~
     \delta_{\ell, \lambda}\delta_{\ell', \lambda'}(-1)^{\ell + \ell' }\sqrt{(2\ell + 1)(2\ell' + 1)} \frac{2\pi^2}{(q_2)^2}\delta\left(q_2 - p_2\right)  \sum_{\tilde{\ell}}(2\tilde{\ell} + 1)\\[-0.2em]
    &
   \times\iiiint \frac{d\Omega_{p_1}}{4\pi}\frac{d\Omega_{p_2}}{4\pi}
   \frac{d\Omega_{q_1}}{4\pi} \frac{d\Omega_{q_2}}{4\pi}
   P_\ell\left(\hatvect{p}_1\cdot\hatvect{p}_2\right)
   P_{\tilde{\ell}}\left(\hatvect{p}^J_1(\vect{p}_1,\vect{p}_2)\cdot\hatvect{p}_2^J(\vect{p}_2)\right)\\
   &
   \hspace{1cm}\times   P_{\tilde{\ell}}\left(\hatvect{q}^J_1(\vect{q}_1,\vect{q}_2)\cdot\hatvect{q}_2^J(\hatvect{q}_2)\right)
   P_{\ell'}\left(\hatvect{q}_1\cdot \hatvect{q}_2\right)\widetilde{T}_{33,~\textrm{JJ}}^{\widetilde{\ell}}\left(p_1^J(\vect{p}_1, \vect{p}_2); q_1^J(\vect{q}_1, \vect{q}_2),q_2^J(\vect{q}_2)\right).
    \end{aligned}
    \label{eq:T_33_SS}
\end{align}
 For the angular integrals above, one can choose $\hatvect{p}_2$($\hatvect{q}_2$) along the $z$-axis and numerically integrate over $\Omega_{p_1}$ ($\Omega_{q_1}$). 
 
\section{Including or subtracting simple pole(s) in an integral equation}
\label{app:Cauchy}
Consider the following integral equation
\footnote{The ideas used here are similar to the so-called $K$-matrix method 
(see, e.g., Ref.~\cite{fbp1983,Ji_2013} for the $K$-matrix method in three-body systems), which is not to be confused with the kernels (e.g., $K(x,y)$, $K_l^h$, and $K'_{23}$) of the integral equations in this paper.}
\begin{align}
    f(x) = \int_0^\Lambda K(x,y)f(y) dy.
\end{align}
Suppose the kernel of the integral equation $K(x,y)$ only contains one simple pole at $y = y_0\pm i\epsilon$ with a corresponding residue $R(x,y_0)$ and $y_0$ is real. (Multiple poles can be treated in a similar manner.) Adding and subtracting the pole gives
\begin{align}
    f(x) = \int_0^\Lambda \left(K(x,y)f(y) - \frac{R(x,y_0)}{y-(y_0 \pm i\epsilon)}f(y_0)\right)dy + \int_0^\Lambda \frac{R(x,y_0)}{y-(y_0\pm i\epsilon)}f(y_0)dy.
    \label{eq:intgEqn}
\end{align}
 The second term can be evaluated using the Cauchy principal value prescription:
\begin{align}
    &\int_0^\Lambda \frac{R(x,y_0)}{y-(y_0\pm i\epsilon)}f(y_0)dy \nonumber\\
=&R(x,y_0)f(y_0)\left(\pm i\pi \theta(\Lambda-y_0)+ \ln{\abs{\frac{\Lambda - y_0}{ y_0}}}\right)\nonumber\\
=&R(x,y_0)f(y_0)\ln{\left(\frac{\Lambda - y_0}{y_0}(-1 \pm i\epsilon)\right)},
\label{eq:Cauchy}
\end{align}
where $\theta$ on the second line is the step function. The logarithm function on the third line has a branch cut on $(-\infty,0]$ and is continuous from above on it.  Plugging Eq.~\eqref{eq:Cauchy} into Eq.~\eqref{eq:intgEqn} gives 
\begin{align}
    f(x) = \int_0^\Lambda \left(K(x,y)f(y) - \frac{R(x,y_0)}{y-(y_0 \pm i\epsilon)}f(y_0)\right)dy + R(x,y_0)f(y_0)\ln{\left(\frac{\Lambda - y_0}{y_0}(-1 \pm i\epsilon)\right)}.
\end{align}
One can then discretize $x$ and $y$ with a finite-dimensional vector $\vect{v}$ with its $i$-th component denoted by $v_i$. The resulting matrix equation can be written compactly in a block matrix form:
\begin{align}
    \begin{bmatrix} 
        f(v_i)\\
        f(y_0)
    \end{bmatrix}
    =
    \begin{bmatrix} 
        K(v_i,v_j) w_j&R(v_i,y_0)\left(\ln{\left(\frac{\Lambda - y_0}{y_0}(-1 \pm i\epsilon)\right)} - \sum_k \frac{w_k}{v_k-y_0}\right)\\
        K(y_0,v_j)w_j&R(y_0,y_0)\left(\ln{\left(\frac{\Lambda - y_0}{y_0}(-1 \pm i\epsilon)\right)} - \sum_k \frac{w_k}{v_k-y_0}\right)
    \end{bmatrix}
    \begin{bmatrix} 
        f(v_j)\\
        f(y_0)
    \end{bmatrix},
\label{eq:discIntgEqn}
\end{align}
where $w_j$ is the weight associated with the $v_j$. In Eq.~\eqref{eq:discIntgEqn}, $f(v_i)$, $f(v_j)$, and $K(y_0,y_j)w_j$ are understood as column vectors, $R(v_i,y_0)$ is understood as a row vector, and $K(v_i,v_j)w_j$ is understood as a block matrix. Note that one can always choose $\vect{v}$ that does not contain the pole, $y_0$.
If $\Lambda > y_0$,  Eq.~\eqref{eq:discIntgEqn} includes the contribution from the simple pole properly and is numerically stable. If $\Lambda < y_0$, the logarithm in Eq.~\eqref{eq:discIntgEqn} automatically cancels with $ \sum_k \left(w_j/(v_k-y_0)\right)$ up to numerical errors and Eq.~\eqref{eq:discIntgEqn} still holds.

Subtracting the contribution from simple poles is accomplished by dropping the $i\pi$ term from the integral and only keeping the Cauchy principal value for each pole. To illustrate how this works in the four-boson calculation, consider $K'_{33}\Gamma_{3}$ in the four-boson integral equation~\eqref{eq:formal_Intg_T3}. This illustration will only use Jacobi momenta and consider pair-wise S waves between the dimer and either of the two single bosons, but the same method also works for different momentum bases. The four-boson state with total energy $E_4$ and pair-wise S wave states is denoted
\begin{align}
     | E_4, k_1 k_2\rangle\equiv | E_4, L = 0, m_L = 0, (\ell = 0, \lambda = 0), k_1 k_2 \rangle,
\end{align}
where the notation is the same as of Eq.~\eqref{eq:pwb-full} but also indicates $E_4$. 
The matrix element of $K'_{33}\Gamma_{3}$ with an incoming four-boson state $|\{\boldsymbol{\alpha}\}\rangle$ and an outgoing state $| E_4, k_1 k_2\rangle$  reads
\begin{align}
    &\langle E_4,p^J_1 p^J_2|K'_{33}\Gamma_{3}|\{\boldsymbol{\alpha}\}\rangle \nonumber \\
    =& \iint_{0}^{\Lambda}\frac{(q_1^J)^2dq_1^J}{2\pi^2}\frac{(q_2^J)^2dq_2^J}{2\pi^2}\langle  E_4,p^J_1 p^J_2|\mathcal{P}_{33}K_{33}(1-K_{33})^{-1}| E_4,q_1^J q_2^J\rangle\Gamma_{3}(q_1^J ,q_2^J)\nonumber \\
    =& \int_{0}^{\Lambda}\frac{(q_1^J)^2dq_1^J}{2\pi^2}\langle E_3(p_1^J),p^J_2|K(1-K)^{-1}|E_3(p_1^J),q_1^J\rangle\Gamma_{3}(q_1^J, p_1^J),
    \label{eq:k33-Gamma3}
\end{align}
where Eqs.~\eqref{eq:K's} and~\eqref{eq:T_33} are used on the second line. On the third line $K_{33}$ have been expressed in terms of the three-boson kernel $K$, introduced in Eq.~\eqref{eq:3boson-OpForm}, and the Dirac-delta function from the spectator boson has been integrated over. The matrix element of $K$ is given by Eq.~\eqref{eq:K_l^h}. $E_3(p_1^J) = E_4 - 2(p_1^J)^2/(3m)$ is the energy of the three-boson subsystem in its CM frame. The spectral decomposition of $K$ (including the measure in the momentum integral) restricted to S waves at a three-boson CM energy $E_3$ is
\begin{align}
     \widehat{K}_0^{h}=&\iint_0^{\Lambda}|E_3,p\rangle \frac{p^2dp}{2\pi^2} K_0^h(E_3, p, q) \frac{q^2dq}{2\pi^2} \langle E_3,q| \nonumber \\
    =& \sum_{a}|E_3, v_a\rangle \lambda_{a}(E_3)\langle E_3,v_a|,
    \label{eq:spec-decomp-K}
\end{align}
where $ \lambda_{i}(E_3)$ are the eigenvalues and $v_i(E_3, p) = \langle p|E_3, v_i\rangle$ are the eigenvectors of $\widehat{K}_0^{h}$ and the sum over $a$ could include an integral for the continuous part of the spectrum. When $E_3$ equals a trimer binding energy $-B_3^{(i)}$, an eigenvalue, denoted $\lambda_1$, equals one. The corresponding eigenvector $v_1(E_3, p)$ is just the trimer vertex function $\mathcal{G}(E_3,p)$, as shown in Eq.~\eqref{eq:G_eigenv}. Plugging Eq.~\eqref{eq:spec-decomp-K} into Eq.~\eqref{eq:k33-Gamma3} gives
\begin{align}
    &\langle E_4,p^J_1 p^J_2|K'_{33}\Gamma_{3}|\{\boldsymbol{\alpha}\}\rangle \nonumber \\
    =& \sum_{a}\frac{\lambda_a\left(E_3(p_1^J)\right)}{1-\lambda_a\left(E_3(p_1^J)\right)}v_a\left((E_3(p_1^J),p_2^J \right)\int_{0}^{\Lambda}\frac{(q_1^J)^2dq_1^J}{2\pi^2}v^{*}_a\left(E_3(p_1^J),q_1^J\right)\Gamma_3(q_1^J, p_1^J) \nonumber \\
    \equiv& \sum_{a}\frac{\lambda_a\left(E_3(p_1^J)\right)}{1-\lambda_a\left(E_3(p_1^J)\right)} F(p_1^J, p_2^J),
\end{align}
which has a pole at $E_3(p_{1,~\textrm{pole}}^{J}) = -B_3^{(i)}$ that comes from $\lambda_1(-B_3^{(i)}) = 1$, assuming $ F(p_1^J, p_2^J)$ is well-behaved\footnote{$ F(p_1^J, p_2^J)$ could have poles at some $p_2^J = p_{2,~\textrm{pole}}^{J}$ from iterations of three-boson amplitudes, but it is well-behaved in $p_1^J$ at all other values of  $p_2^J$.} around $p_1^J = p_{1,~\textrm{pole}}^{J}$. A general integral over $p^J_1$ that involves $\langle E_4,p^J_1 p^J_2|K'_{33}\Gamma_{3}|\{\boldsymbol{\alpha}\}\rangle$ can be written as
\begin{align}
    &\int_0^\Lambda dp^J_1 f(p_1^J, p_2^J)\sum_{a}\frac{\lambda_a\left(E_3(p_1^J)\right)}{1-\lambda_a\left(E_3(p_1^J)\right)} F(p_1^J, p_2^J) \nonumber\\
    =&\left(\int_{0}^{ p_{1,~\textrm{pole}}^{J}- \epsilon} + \int_{ p_{1,~\textrm{pole}}^{J}- \epsilon}^{ p_{1,~\textrm{pole}}^{J}+ \epsilon} + \int_{ p_{1,~\textrm{pole}}^{J}+ \epsilon}^{\Lambda}\right) dp^J_1 f(p_1^J, p_2^J)\sum_{a}\frac{\lambda_a\left(E_3(p_1^J)\right)}{1-\lambda_a\left(E_3(p_1^J)\right)} F(p_1^J, p_2^J) ,
    \label{eq:intg-over-spec-decomp}
\end{align}
where $f(p_1^J, p_2^J)$ is an arbitrary function that is smooth around $p_1^J = p_{1,~\textrm{pole}}^{J}$ and $\epsilon$ is an infinitesimal positive number. The integral over $p_1^J$ has been split into three parts, and the sum of the first and third parts gives the Cauchy principal value of the integral. The second part has zero measure as $\epsilon \to 0$ and its only contribution comes from the pole where $\lambda_1\left(E_3(p_{1,~\textrm{pole}}^{J})\right) = 1$. In other words, 
\begin{align}
     \int_{ p_{1,~\textrm{pole}}^{J}- \epsilon}^{ p_{1,~\textrm{pole}}^{J}+ \epsilon} dp^J_1 f(p_1^J, p_2^J)\sum_{a\neq 1}\frac{\lambda_a\left(E_3(p_1^J)\right)}{1-\lambda_a\left(E_3(p_1^J)\right)} F(p_1^J, p_2^J) = 0.
\end{align}
Therefore, if one subtracts the deep trimers, i.e., removes $\lambda_1(-B_3^{(i)})$ and $|-B_3^{(i)}, v_1\rangle$ corresponding to the deep trimers from the spectral decomposition of $K$ in Eq.~\eqref{eq:spec-decomp-K},  then what is left in Eq.~\eqref{eq:intg-over-spec-decomp} is just the Cauchy principal value. Note that the pole only exists at $p_1^J = p_{1,~\textrm{pole}}^J$. Therefore, no subtraction is needed at $p_1^J \neq p_{1,~\textrm{pole}}^J$, and all states $|E_3(p_1^J) \neq -B_3^{(i)}, v_a\rangle$ with any $a$ still contribute.

\section{Discussion on momentum and angular momentum cutoffs}
\label{app:diffcutoffs}

In principle, the cutoff $\Lambda$ used in the integral equation~\eqref{eqn:integral equation for three-boson scatterings} for the dimer-single-boson scattering amplitude can be different from the cutoff $\Lambda'$ of the \fourBodyMom in the four-boson integral equation~\eqref{eq:4body-intg-full}. This is similar to using a cutoff $\Lambda_\textrm{2body}$ for the two-body subsystem of a three-body system with a different cutoff $\Lambda \neq \Lambda_\textrm{2body}$; taking $\Lambda_\textrm{2body}\to \infty$ while keeping $\Lambda$ finite is discussed in Ref.~\cite{Bedaque_2000}. In four-boson calculations, the relationship between $\Lambda$ and  $\Lambda'$ is similar but slightly more complicated due to the deep trimer poles.
If the deep trimers are properly addressed, the four-body vertex should not depend on $\Lambda$ for $\Lambda \gg \Lambda'$ at large $\Lambda$ and $\Lambda'$ since the dimer-single-boson scattering amplitude, $t_l^h(E,k,p)$ in Eq.~\eqref{eqn:integral equation for three-boson scatterings}, converges for $p, k\ll \Lambda$ as $\Lambda\to\infty$. 
Furthermore, to compute the kernel of the four-boson integral equation~\eqref{eq:4body-intg-full}, $t_l^h(E,k,p)$ needs to be evaluated at, e.g., $p = p_1^J = \left|\vect{p}_1 + \vect{p}_2/3\right|\lesssim 4\Lambda'/3$ (see Eqs.~\eqref{eq:Ttilde_33_JN} and~\eqref{eq:T33_Jacobi}), but $t_l^h(E,k,p)$ may not be cutoff-independent for $p,k\sim\Lambda'$ if $ \Lambda'\sim \Lambda$. However, the contribution from the \fourBodyMom $p_1, p_2 \sim\Lambda'$ to the four-boson loop integrals is expected to be suppressed compared to the contribution from $p_1, p_2$ of the typical scale of the system under consideration. This also aligns with the diagrammatic results using $\Lambda'= \Lambda$ as shown in Fig.~\ref{fig: 4body-bindings} where a good convergence of tetramer binding energies is found. A quantitative and analytic study on the impact of using different values for $\Lambda$ and $\Lambda'$ requires a more detailed asymptotic analysis of the four-boson integral equation, which is beyond the scope of this paper. 

Regarding the cutoff, $\Lambda_E$, of the energy loop integral, the choice $\Lambda_E=4\Lambda^2/m$ is found to be sufficiently large to compute tetramer binding energies. In order to justify the choice of $\ell_{\textrm{max}} = 2$, Table~\ref{tab:E_4_lmax} shows $E_4^{(0)}$ as a function of $\ell_{\textrm{max}}$ calculated at $\Lambda = \Lambda' = 4000\sqrt{mB_2}$. 
\begin{table}[]
    \centering
    \begin{tabular}{|ccccccc|}
    \hline
    $\ell_{\textrm{max}}$&0&1&2&3&4&5\\\hline
    $E_4^{(0)}[\sqrt{mB_2}]$& 416.2& 424.6 & 425.5& 425.8& 425.9& 425.9\\\hline
    $(E_4^{(1)} - B_3^{(0)})[\sqrt{mB_2}]$& Unbound& 0.0099 & 0.0135& 0.0143& 0.0150& 0.0152\\\hline
    \end{tabular}
    \caption{Tetramer binding energies for different angular momentum cutoff $\ell_{\textrm{max}}$ evaluated at $\Lambda = 4000\sqrt{mB_2}$ for cold $^4$He atoms with deep trimers subtracted.}
    \label{tab:E_4_lmax}
\end{table}
It is found that $E_4^{(0)}$ only receives small correction from $\ell > 2$ and therefore $\ell_{\textrm{max}} = 2$ is used in this paper to reduce numerical expenses. For example, in the matrix element of $K'_{22}$ given by Eq.~\eqref{eq:K22'}, the contribution from $K_{23}(1+T_{33})K_{32}$, which corresponds to diagrams with two or more boson exchanges between the dimer and single boson in the three-boson subsystem, is approximated using partial waves below $\ell_{\textrm{max}} = 2$. On the other hand, the angular integrals in $K_{22}$, which corresponds to a diagram with only one boson exchange in the three-boson subsystem, are evaluated directly without partial waves. This is because the angular integrals in $K_{22}$ are much easier to evaluate exactly, as given by Eq.~\eqref{eq:K22}, than using partial-wave projections. Although this means the matrix element of $K_{22}$, compared to other components of the kernel with a finite partial-wave cutoff $\ell_{\textrm{max}}$, always contains contamination from higher partial waves, the matrix element of $K_{22}$ is independent of $\ell_{\textrm{max}}$. Therefore, a convergence of four-boson observables as a function of $\ell_{\textrm{max}}$ is still expected. This is also demonstrated in Table~\ref{tab:E_4_lmax}.
\section{Complex eigenvalues and locations of the resonances}
\label{app: Eigenvalues}
Consider an eigenvalue $\lambda(E+i\Gamma/2)$ of the kernel of an integral equation as a function of a complex energy variable $E+i\Gamma/2$, where both $E$ and $\Gamma$ are real. When $\lambda(E_0+i\Gamma_0/2) = 1$ with $\Gamma_0\neq 0$ , a resonance is identified with a binding energy $E_0$ and a decay width $\Gamma_0$. In general, a resonance is much more difficult to locate than a real bound state since the former requires solving for $E_0+i\Gamma_0/2$ in the complex plane. However, for $\Gamma_0/(2E_0)\ll 1$ one can find $E_0$ and $\Gamma_0$ perturbatively. Assuming $\lambda(E+i\Gamma/2)$ is analytic around $E_0$ with a radius of convergence larger than $\Gamma_0/2$, one can expand $\lambda(E_0+i\Gamma_0/2)$ around $E_0$:
\begin{align}
 1=\lambda(E_0 + i\Gamma_0/2) &= \lambda(E_0) + E_0\frac{\partial \lambda}{\partial E}\bigg\rvert_{E = E_0}\frac{i\Gamma_0}{2E_0} + \cdots\nonumber\\
 &\equiv \lambda(E_0) + E_0\lambda'(E_0)\frac{i\Gamma_0}{2E_0} + \cdots.
 \label{eq:EigenvalueExpansion}
\end{align}
In order to find $E_0$ and $\Gamma_0$ perturbatively, one can treat the imaginary part of $\lambda(E_0)$ as a perturbation:
\begin{align}
    \alpha \equiv \Im(\lambda(E_0))\ll\Re(\lambda(E_0))\sim \mathcal{O}(1)
\end{align}
and use the following series expansions of $E_0$ and $\Gamma_0$:
\begin{align}
    E_0 &= \sum_{n=0,1,\cdots}\alpha^n E_0^{(n)}\\
    \Gamma_0 &= \sum_{n=1,2,\cdots}\alpha^{n}\Gamma_0^{(n)},
\end{align}
where the expansion for $\Gamma_0$ starts at $n = 1$ because $\Gamma_0\ll E_0$ and thus $\Gamma_0$ should be driven by $\Im(\lambda(E_0))$. 
The real part of Eq.~\eqref{eq:EigenvalueExpansion} to the lowest order can be used to solve for $E_0^{(0)}$:
\begin{align}
    \Re(\lambda(E_0^{(0)}))  &= 1,  
\label{eq:E0(0)}
\end{align}
The imaginary part of Eq.~\eqref{eq:EigenvalueExpansion} to the lowest non-vanishing order reads
\begin{align}
    \alpha +  \Re(\lambda'(E_0^{(0)}))\frac{\alpha\Gamma_0^{(1)}}{2} = 0,
    \label{eq:gamma0(0)}
\end{align}
which gives
\begin{align}
    \frac{\alpha\Gamma_0^{(1)}}{2}= -\frac{\Im(\lambda(E_0^{(0)}))}{\Re(\lambda'(E_0^{(0)}))}.
    \label{eq:gamma_0}
\end{align}

In order to find the next-order corrections and/or the error of the first-order approximations above, one can take the real part of Eq.~\eqref{eq:EigenvalueExpansion} and first collect terms of order $\alpha$:
\begin{align}
     \Re(\lambda'(E_0^{(0)}))\alpha E_0^{(1)}= 0.
\end{align}
In order to find the first non-vanishing correction to $E_0^{(0)}$, one can collect terms of order $\alpha^2$ for the real part of Eq.~\eqref{eq:EigenvalueExpansion}:
\begin{align}
     \Re(\lambda'(E_0^{(0)}))\alpha^2 E_0^{(2)} - \alpha \frac{\Im(\lambda'(E_0^{(0)}))}{\Im(\lambda(E_0^{(0)}))}\frac{\alpha \Gamma_0^{(1)}}{2} - 
     \frac{1}{2}\Re(\lambda''(E_0^{(0)}))\frac{ \left(\alpha\Gamma_0^{(1)}\right)^2}{4} = 0,
\end{align}
where all vanishing terms have been dropped. $\alpha^2 E_0^{(2)}$ is given by
\begin{align}
    \alpha^2 E_0^{(2)} = \frac{\Im(\lambda'(E_0^{(0)}))\alpha\Gamma_0^{(1)}+\frac{1}{4}\Re(\lambda''(E_0^{(0)}))\left(\alpha\Gamma_0^{(1)}\right)^2}{2\Re(\lambda'(E_0^{(0)}))}.
    \label{eq: E0(2)}
\end{align}
Similarly, keeping the imaginary part Eq.~\eqref{eq:EigenvalueExpansion} with terms of order $\alpha^2$ yields $\Gamma_0^{(2)} = 0$. In order to find the first non-vanishing correction to $\Gamma_0^{(1)}$, one can take the imaginary part of Eq.~\eqref{eq:EigenvalueExpansion}. The non-vanishing terms of order $\alpha^3$ give
\begin{align}
    &\alpha \frac{\Im(\lambda'(E_0^{(0)}))}{\Im(\lambda(E_0^{(0)}))}\alpha^2 E_0^{(2)} +  \Re(\lambda'(E_0^{(0)}))\frac{\alpha^3\Gamma_0^{(3)}}{2}+\Re(\lambda''(E_0^{(0)}))\alpha^2 E_0^{(2)}\frac{\alpha\Gamma_0^{(1)}}{2}  \nonumber\\
    &\qquad -\frac{1}{2}\alpha \frac{\Im(\lambda''(E_0^{(0)}))}{\Im(\lambda(E_0^{(0)}))}\left(\frac{\alpha\Gamma_0^{(1)}}{2}\right)^2 - \frac{1}{6}\Re(\lambda'''(E_0^{(0)}))\left(\frac{\alpha \Gamma_0^{(1)}}{2}\right)^3 = 0,
\end{align}
which can be solved for $\alpha^3 \Gamma_0^{(3)}$. 

In practice, one can first solve for $E_0^{(0)}$ and $\Gamma_0^{(1)}/2$, then compute $\lambda'(E_0^{(0)})$ numerically to find perturbatively $E_0^{(2)}$ and $\Gamma_0^{(3)}$, and at last use $E_0^{(2)}$ and $\Gamma_0^{(3)}$ to estimate the uncertainties of $E_0^{(0)}$ and $\Gamma_0^{(1)}/2$, denoted by Err$(E_0^{(0)})$ and Err$(\alpha\Gamma_0^{(1)}/2)$, respectively. This gives
\begin{align}
    \textrm{Err}(E_0^{(0)}) &\approx \abs{\frac{\Im(\lambda'(E_0^{(0)}))\Im(\lambda(E_0^{(0)}))}{\left[\Re(\lambda'(E_0^{(0)}))\right]^2}}\nonumber\\
    \textrm{Err}(\alpha\Gamma_0^{(1)}/2) &\approx \abs{\left(\frac{\Im(\lambda'(E_0^{(0)}))}{\Re(\lambda'(E_0^{(0)}))}\right)^2\frac{\Im(\lambda(E_0^{(0)}))}{\Re(\lambda'(E_0^{(0)}))}}.
    \label{eq:eigen-error}
\end{align}
This uncertainty estimation works as long as the second and higher derivatives of $\lambda(E)$ at $E = E_0^{(0)}$ are of a similar size as, or smaller than, the first derivative.
Note that Eq.~\eqref{eq:E0(0)} may not have a solution. In that case, one may pick a real number $\beta  \ll 1$, add and subtract it on the left-hand side of Eq.~\eqref{eq:EigenvalueExpansion}, solve $\Re(\lambda(E_0^{(0)}))  = 1-\beta$ instead of Eq.~\eqref{eq:E0(0)}, and treat $\beta$ as a correction term.

\bibliography{references.bib}
\end{document}